\documentclass[11pt]{article}
\usepackage{geometry}                
\usepackage{epstopdf}
\usepackage{graphicx}
\usepackage{amsfonts, amsbsy, amssymb, amsmath, graphicx, float, subfigure, color}
\usepackage{subfigure}
\def\Ra{\mbox{\textrm{Ra}}}

\begin{document}
\title{{ Symmetry and} plate-like convection  in fluids with temperature-dependent viscosity}
\author{Jezabel Curbelo$^{1,2}$, Ana M. Mancho$^1$\\
$^1$Instituto de Ciencias Matem\'aticas (CSIC-UAM-UCM-UC3M), \\
C/ Nicol\'as Cabrera, 13-15, 28049 Madrid, Spain\\
$^2$Departamento de Matem\'aticas, Universidad Aut\'onoma de Madrid, \\ Facultad de Ciencias, m\'odulo 17, 28049, Madrid, Spain}


\maketitle
\begin{abstract}

{We explore the instabilities developed in a fluid in which viscosity  depends on temperature. In particular, we consider a dependency
that models { a very viscous (and thus rather rigid) lithosphere over a convecting mantle.}
To this end, we study a  2D convection problem in which viscosity depends  on temperature by abruptly changing  its value by a factor of 400  within a narrow temperature gap. We conduct a study which combines bifurcation analysis and time-dependent simulations. Solutions such as limit cycles are found that are fundamentally related to the presence of symmetry. Spontaneous plate-like behaviors that rapidly evolve towards a stagnant lid regime emerge sporadically  through abrupt  bursts during these cycles.
 The plate-like evolution alternates motions towards either the right or the left, thereby  introducing temporary asymmetries on the convecting styles. Further time-dependent regimes with stagnant  and plate-like lids are found and described.}
\end{abstract}

\section{Introduction}

 Rayleigh-B\'enard convection is the classic example of thermal convection \cite{ben1900}. In these systems, under certain critical conditions, small fluctuations lead to massive reorganization of the convective motions \cite{MBH97,HHM02,HMHGD05,NMH07}.
 This is a characteristic phenomenon of open systems that transfer energy and that are modelled by nonlinear equations.
  The internal energy of  the planetary interiors is dissipated by convective processes, thus convection plays a crucial role in the evolution of the planet.
 Convective styles in planetary interiors are different from Rayleigh-B\'enard convection.
 For instance, plate tectonics,  which is distinctive of the Earth \cite{KLG02}, is the surface manifestation of convection in the Earth's mantle.
Other bodies in the solar system, such as the Moon, Venus or Mars  do not exhibit  plate tectonics and present
other  convection expressions with a stagnant lithosphere \cite{SM97, Setal92}.  Different physical  justifications exist for the diverse types of convection:  layered convection, for instance, is due to  endothermic phase changes in the minerals that constitute the mantle interior \cite{MW91}.
{
The mantle is  compressible  due to changes in density, which increases towards the Earth interior. Numerical analysis of compressible convection indicates that density stratification has a stabilizing effect  \cite{jarvis}, produces   upwelling plumes weaker than those downwelling and influences the thermal boundary layer   \cite{zhong} .
The dependence of conductivity on temperature introduces new nonlinearities into the heat equation, which may lead to diverse dynamics \cite{hof}. When conductivity decreases  with temperature, convection becomes more chaotic and time-dependent \cite{Dubu, Yana}. Thermal conductivity  variation has
 generally been less studied than that of viscosity,  as the latter is much stronger in the Earth's mantle.}
Large viscosity contrasts in fluids with temperature-dependent viscosity  lead to stagnant lid convection \cite{MS95,SM96}. {Regarding the subduction initiation, numerical results
\cite{T98,TH98,B03,S04} suggest that this}
is only possible if the stiff upper layers of the  lithosphere are weakened by brittle fracture.  Several mechanisms have been proposed for driving the motion of the lithospheric plates. Forsyth and Uyeda \cite{FU75}, for instance, conclude that plate-like motion is produced by the sinking slab that  pulls the plate in the subduction process {due to an excess of lithosphere density.}

{
 Finding the impact  of the different physical properties present in the mantle on its convection styles is an important goal of research into planetary interiors.
In this context, our focus is on examining the instabilities found in a 2D fluid    in the presence of the O(2) symmetry which  contemplates  a phase transition similar to a melting-solidification processes.
 In particular, we   consider  a highly viscous layer (lithosphere)
over a fluid mantle which is modeled with a viscosity that changes abruptly  by a factor of 400, in a narrow temperature gap  at which magma melts.
In  phase transitions,  other fluid properties in addition to viscosity may change abruptly, such as  density or  thermal diffusivity. However, in this study we confine ourselves solely to the  effects due to the variability of viscosity, since consideration of the effect of  simultaneous variations on all the properties prevents a focused  understanding of the exact role played by each one of these properties.
Viscosity  is a measure of  fluid  resistance to gradual deformation, and in this sense highly viscous fluids are more likely to behave rigidly when compared to less viscous fluids. When examining the proposed transition   with temperature, we focus on the global fluid motion when some parts of this motion tend to be more rigid than others. By disregarding  the   variations on density in this transition, we move away from  instabilities  caused by abrupt density changes such as the Rayleigh Taylor instability, in which a denser fluid over a lighter one tends to penetrate it by forming a fingering pattern. A recent article  by M. Ulvrov\'a  {\it et al.} \cite{ULCRT12} deals with a  problem similar to the one we address here, but  takes into account variations in both  density and  viscosity. Thermal conductivity  effects  are related to the relative importance of heat advection versus diffusion. Diffusive effects
are therefore important at large conductivity, while  heat advection by fluid particles is dominant at low conductivity. The contrasts  arising from these variations are beyond the scope of our work and  are thus disregarded herein.}

{ In our setting we show that   convective}  processes exist  which include plate-like motions that alternate in time with stagnant-lid regimes. {Some of these transitions include bursts in which the solution  releases energy to accommodate different spatial patterns. }
  These solutions are mathematically related to limit cycles,  which  are persistent solutions in the presence of the O(2)
symmetry \cite{AGH88,GH88,PoMa97} which is also found in this problem. There exist numerous novel dynamical phenomena in fluids that are  fundamentally related to the presence of symmetries \cite{CK91}: these  include rotating waves \cite{R73}, modulated waves \cite{AGH88,R82} and stable heteroclinic cycles \cite{AGH88,GH88,PoMa97}.
 The SO(2) symmetry is present   in  the problem under consideration,  because the equations are invariant  under translations  and  periodic boundary conditions do not break this invariance.  Additionally,  if  the reflection symmetry exists,  the full  group of symmetry is the O(2) group.

 The impact of the symmetry on the solutions displayed in convection problems with temperature-dependent viscosity  has been addressed in \cite{CuMa11, CM13}, where
a 2D physical set-up similar to ours is analyzed. The viscosity law  considered in this work is similar to the one studied in   \cite{CM13},  the main difference being that the viscosity change in our current setting is achieved within a narrower temperature gap. Our problem is idealized in terms of realistic geophysical flows occurring in the Earth's interior, as these are 3D flows moving in spherical shells \cite{B75,B82}. Under these conditions,
the symmetry present in the problem  is formed by all the orientation, preserving rigid motions of $ \mathbb{R}^3$ that fix the origin, which is the SO(3) group \cite{C75,GS82,IG84}. The effects of
the  Earth's  rotation are negligible in this respect and do not break this symmetry, as  the high viscosity of the mantle renders the   Coriolis number insignificant. The link between our simplified problem  and these realistic set-ups is that  the O(2) symmetry is isomorphic   to the rotations along the azimuthal coordinate, which form a closed subgroup of SO(3). Furthermore the O(2) symmetry is present in systems with cylindrical geometry, which provide an idealized setting for volcanic conduits and magma chambers.  {The results described in this paper confirm the symmetry  role
in the solutions that under the physical conditions considered exhibit  plate-like dynamics and energy bursts}.

The article is organized as follows. Section 2 describes the physical set-up and provides  the governing equations as well as   a detailed characterization of the viscosity law. Section 3  briefly introduces
the numerical methods used to obtain the solutions. The results  are presented in Section 4.  Finally, Section 5 details the conclusions.

 \begin{figure}
     \centering
   \includegraphics[width=15cm]{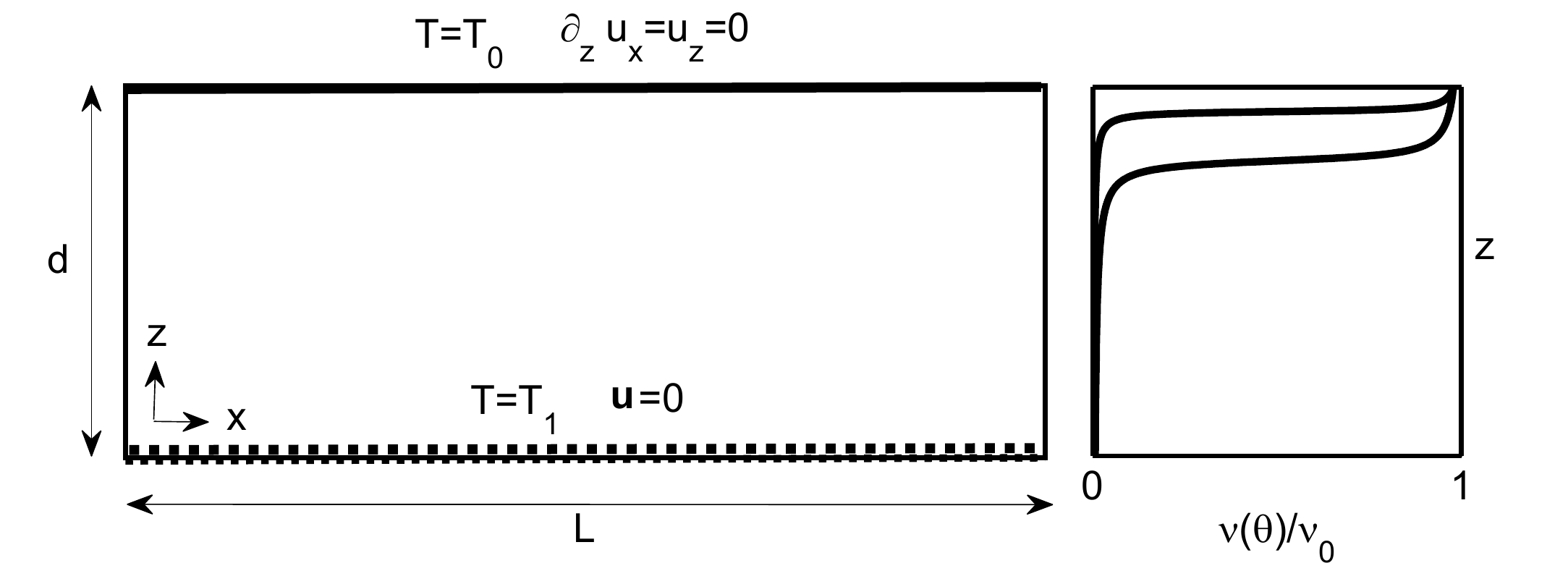}
    \caption{Problem set-up. A 2D container of length $L$ and depth $d$ with periodic lateral boundary conditions. The bottom plate (dashed line)  is rigid and is at temperature $T_1$;  the upper plate (thick line) is  free slip and is at temperature $T_0$ ($T_0<T_1$). {  The viscosity transitions versus the depth $z$ for the conductive temperature and $\Ra=50$ and 150 are  depicted on the right.}}\label{F1}
   \end{figure}

\section{The physical set-up and the governing equations}
We consider a convection problem in a fluid layer of thickness $d$  placed in  a 2D  finite container  of size $L$  as shown in Fig \ref{F1}. The bottom plate is rigid, \emph{i.e.} ${\bf u}=0$, and it is at temperature ${T_1}$. The upper plate   is non-deformable and free slip and  is at temperature  ${T_0},$ where ${T_0}={T_1}-\Delta T$ and $\Delta T$ is the vertical temperature difference, which is positive, {\it i.e},  ${T_0}<{T_1}$.
 The lateral boundary conditions are periodic.

 The equations governing the system are expressed with magnitudes in dimensionless form after
rescaling as  follows: $(x', z') = (x,z)/ d$, $t' = \kappa t / d^2$,
$\mathbf{u}'=  d \mathbf{u}/\kappa$, $P' = d^2 P / (\rho_0 \kappa \nu_0)$ , $\theta'= (T - {T_0}) /(\Delta T)$.
Here, $\kappa$ is the thermal diffusivity, { $\rho_0$ is the mean density at temperature $T_0$  and $\nu_0$ is the  reference viscosity}.
After rescaling the domain, $\Omega_1 = [0,L)\times[0,d]$
is transformed into $\Omega_2 = [0 ,\Gamma)\times[  0 , 1] $ where
$\Gamma  = L / d$ is the aspect ratio.    The non-dimensional equations  are (after dropping the primes in the fields):
   \begin{align}
&\nabla \cdot \mathbf{u}=0, \label{eqproblem1}\\
&\frac{1}{Pr}(\partial_t \mathbf{u}+ \mathbf{u}\cdot \nabla \mathbf{u}) = \Ra\theta\vec{e}_3-\nabla P+\text{div} \left(\frac{\nu(\theta)}{\nu_0}(\nabla \mathbf{u}+(\nabla \mathbf{u})^T)\right),\label{eqproblem2}\\
&\partial_t \theta + \mathbf{u} \cdot \nabla \theta= \Delta \theta.\label{eqproblem3}
\end{align}
Here, $\vec{e}_3$ represents the unitary vector in the vertical direction; $\Ra=d^4 \alpha g \Delta T/(\nu_0 \kappa)$ is the Rayleigh number; $g$ is the gravity acceleration; $\alpha$ the thermal expansion coefficient and $Pr=\nu_0/\kappa$ is the Prandtl number. Typically for rocks, $Pr$  is very large, since they present low thermal conductivity (approximately $10^{-6}m^2/s$) and very large viscosity (of the order $10^{20} N s/m^2$) \cite{Dav01}. Thus, for the problem under consideration, $Pr$ can be considered as infinite and the left-hand side term in \eqref{eqproblem2} can be made equal to zero.
These equations use the  Boussinesq approximation in which the density is considered constant everywhere except in the buoyant term of Eq. \eqref{eqproblem2} where a dependence on temperature is assumed, as follows $\rho=\rho_0(1-\alpha(T-{T_0})).$ Thus, no change in the density  at the melting temperature is considered and is assumed to be  small.   Jointly with equations (\ref{eqproblem1})-(\ref{eqproblem3}), the lateral periodic conditions are invariant under translations along the $x$-coordinate, which introduces the symmetry SO(2) into the problem. The reflexion symmetry $x\to -x$ is also present   insofar as  the fields are conveniently transformed as follows:  $(\theta,u_x,u_z,p) \to(\theta,-u_x,u_z,p)$. In this  case,   the O(2)  group expresses the full  symmetry of the problem.

     \begin{figure}
     \centering
   \includegraphics[width=8.5cm]{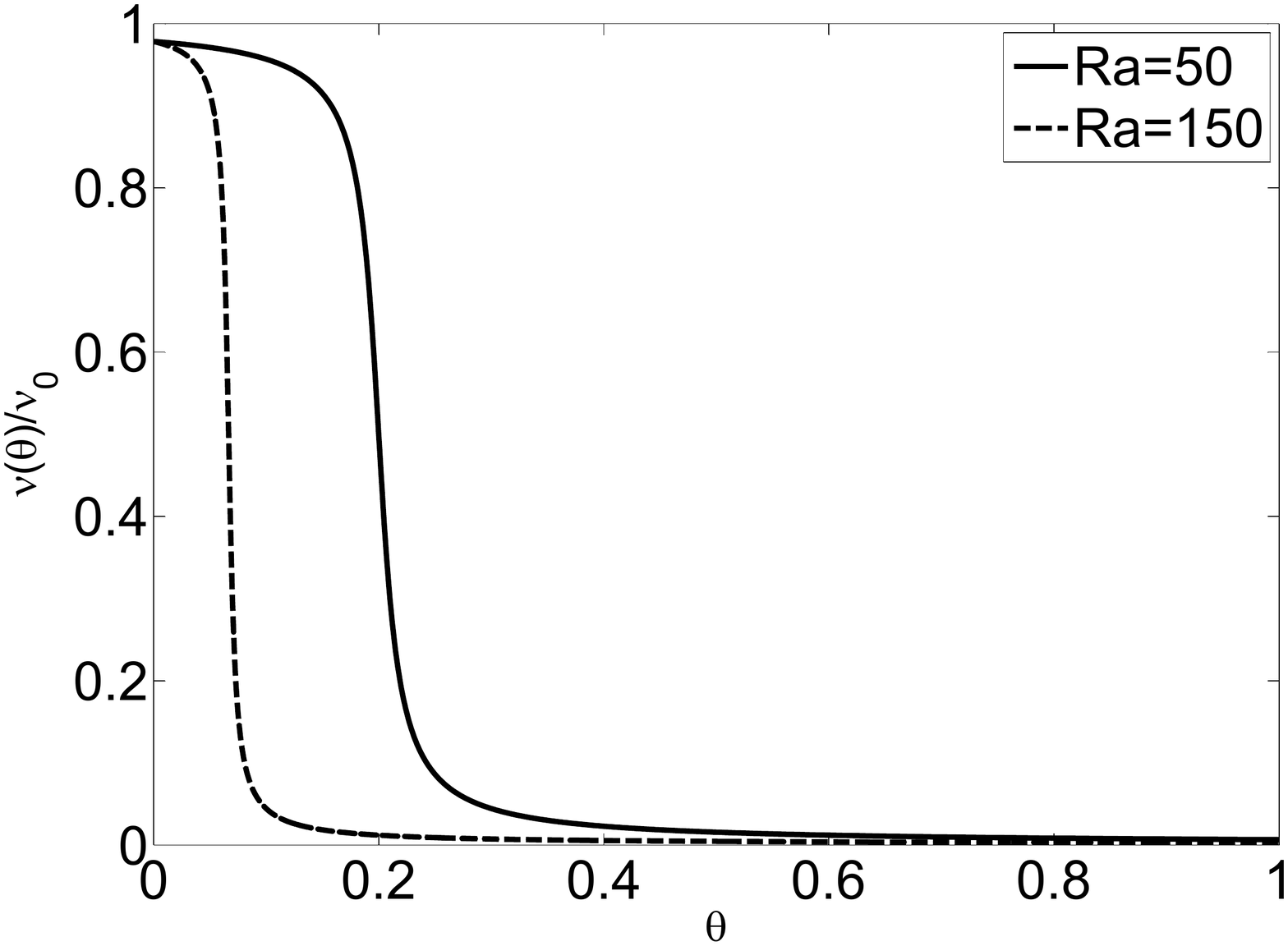}
    \caption{Viscosity law and its dependence on the Rayleigh number.}\label{Fvisco}
   \end{figure}
The viscosity $\nu(\theta)$ is a smooth, positive and bounded function of $\theta$. We use a law that represents the  melting  by means of  an abrupt change in the viscosity
at a small temperature gap defining the melting transition.  In dimensionless form, this law is,
\begin{equation}\label{eqarcotang} \frac{\nu(\theta)}{\nu_0}= - \left( \frac{1-a}{\pi}\right) \arctan(\beta \mu(\Ra\theta  - \Ra_t))+ \left(\frac{1+a}{2}\right) \end{equation}
{Here, the  temperature at which the transition occurs is adjusted by the transition Rayleigh $\Ra_t$ which in our case is $\Ra_t=10$. The choice of a positive value for $\Ra_t$ imposes that there exists a viscosity transition in the interior of the fluid layer, even if $\Ra$ is very large. The parameter $\beta$ controls the abruptness of the viscosity transition on $\theta$. Throughout this study we take  $\beta=100$. The constant $\mu$, fixed to  $\mu=0.$0146,  expresses fluid properties.
The presence of the $\Ra$ number in the viscosity law is uncommon among the literature dealing with  viscosity dependent on temperature.
However, it expresses better what happens in laboratory experiments in which the increment of the $\Ra$ number is performed by increasing the temperature ${T_1}$ at the lower  boundary. This procedure ties the
viscosity to changes in the Rayleigh number, which is the parameter that we vary in our study.
Changes in the $\Ra$ number, as it appears in the viscosity law,
necessarily imply changes in the viscosity contrasts. This is explored in further detail.
The maximum viscosity in the fluid layer is  $\nu_0$, and this is the viscosity value used to define the dimensionless Rayleigh number  {$\Ra$} in Eq. (\ref{eqproblem2}).
In practice,  $\nu_0$  is a viscosity  only  taken by the fluid at the upper surface where $\theta=0$ ({\it i.e},  at temperature {$T_0$}), in  the limit of large $\Ra_t$ and small $\Ra$.}
The parameter $a$  is related to the inverse of the maximum viscosity contrast and is fixed at $10^{-3}$. In the range of $\Ra$ numbers considered in this study, we obtain viscosity contrasts of the order of $3\cdot 10^2-4\cdot 10^2$.  Figure \ref{Fvisco} represents the viscosity law at different $\Ra$ numbers within the range considered in this work. It is observed that as the $\Ra$ number increases the viscosity transition occurs in a temperature gap closer to $\theta=0$, and therefore closer to  the upper surface. { The
  conductive solution (i.e.  $\mathbf{u}_c=0$) to the problem described by equations (\ref{eqproblem1})-(\ref{eqproblem3})  corresponds to the linear temperature $\theta_c=-z+1$. Fig. \ref{F1} shows two viscosity profiles as a function of the depth $z$ for this particular temperature solution.  These profiles are obtained at the same  $\Ra$ numbers as in  Figure \ref{Fvisco}, i.e. $\Ra=50, 150$, and they confirm that the viscosity transition occurs close to the upper surface. }

A viscosity law similar to that expressed in  Eq. (\ref{eqarcotang}) has been proposed in \cite{ULCRT12,CM13}. {The study in \cite{ULCRT12}
 also considers  a change in the density at the temperature of  transition, while  in \cite{CM13} the viscosity change occurs within a broader temperature gap.

%
\section{Numerical methods}
The results presented in this work are obtained by solving the basic equations and boundary conditions with the numerical techniques
reported in \cite{CuMa11}.  Our analysis is assisted by time-dependent numerical simulations and bifurcation techniques such as branch continuation. These schemes are briefly described below.

\subsection{Stationary solutions and their stability}

 \begin{figure}[!h]
 \begin{center}
   \includegraphics[width=8.5cm]{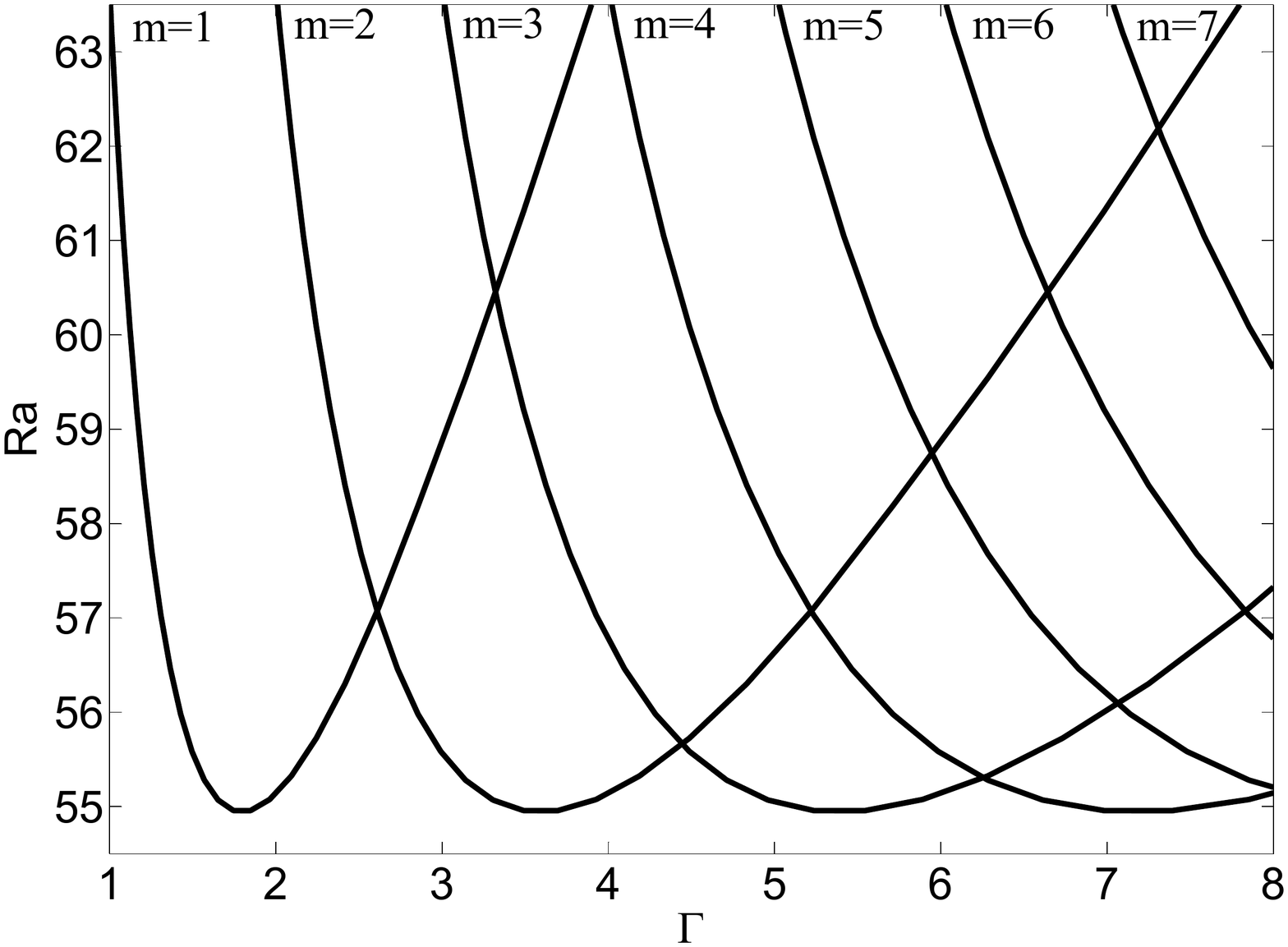}
    \caption{Critical instability curves $\Ra(m,\Gamma)$ for a fluid layer with temperature dependent viscosity taking  $\mu=0.0146$, $a=0.001$, $\Ra_t=10$ and $\beta=100$.}\label{F2}
    \end{center}
   \end{figure}

 The simplest stationary solution to the problem described by equations (\ref{eqproblem1})-(\ref{eqproblem3}) and their boundary conditions is the
  conductive solution that satisfies $\mathbf{u}_c=0$ and $\theta_c=-z+1$.  This solution is stable only  for a range of  vertical temperature gradients that are represented by
small enough Rayleigh numbers. Beyond the
critical threshold $\Ra_c$,  a convective motion settles in and new structures are observed, which
may be either time-dependent or stationary. In the latter case, the stationary equations, obtained by cancelling the time derivatives in the system  (\ref{eqproblem1})-(\ref{eqproblem3}) are satisfied
by the bifurcating solutions. At the instability threshold of the conductive state, the growing solutions are periodic and correspond to sine or cosine eigenfunctions with wave number $m$. Figure \ref{F2} displays the critical instability curves for different $m$ values as a function of the aspect ratio. These curves are obtained  by means of a simplified linear stability analysis for the conductive solution, as reported in \cite{CuMa11}.
At the instability threshold around $\Ra\sim 55$, 
the viscosity law indicates (see Figure \ref{Fvisco}) that the viscosity transition  takes place at the lowest  temperatures across the fluid layer, which is near the fluid surface.  Thus, at this threshold the fluid consists of a highly viscous
 layer over a fluid that is not so viscous and is starting its convection.

Beyond the instability thresholds displayed in Figure  \ref{F2}, new branches of stationary  solutions arise that evolve with the external
physical parameters. There also exist new critical thresholds  at which  stability is lost, thereby giving rise to new bifurcated structures.
These stationary solutions are numerically obtained  by using  an iterative Newton-Raphson method as reported in \cite{CuMa11, CM13}.

The study of the stability  of the stationary solutions under consideration is  addressed  
 by means of a linear stability analysis. To this end, a field  ${Y}$ representing the unknown physical magnitudes  is decomposed into its
stationary solution  $Y^b$ and a perturbation $\tilde{{y}}$ as follows:
\begin{align}\label{pertur}
{Y}(x,z,t)={Y}^b(x,z)+\tilde{{y}} (x,z){\rm e}^{\lambda t}.
\end{align}
The sign in the real part of the eigenvalue $\lambda$ determines the stability of
the solution:  if it is  negative, the perturbation decays and the stationary solution
is stable, while if it is positive the perturbation grows over time and the stationary solution
is unstable.

For each unknown field  expression, (\ref{pertur}) is introduced into the system (\ref{eqproblem1})-(\ref{eqproblem3}) and the  equations are linearized in  $\tilde{{y}}$, which are assumed to be small (see \cite{CuMa11,CM13} for details). Together with their  boundary conditions, the equations   define a generalized eigenvalue problem.
The unknown perturbation fields $\tilde{y} $ of the linear equations  are approached by means of a spectral method according to the expansion:
\begin{align}\label{eqexpansion3}
\tilde{y}(x,z)=&\sum_{l=1}^{\lceil L/2\rceil}\sum_{m=0}^{M-1} b^{\tilde{y}}_{lm}T_m(z)\cos((l-1)x) + \sum_{l=2}^{\lceil L/2\rceil}\sum_{m=0}^{M-1} c^{\tilde{y}}_{lm}T_m(z)\sin((l-1)x).
\end{align}
In this  notation,  $\lceil \cdot \rceil$ represents the nearest integer towards infinity. Here,  $L$ is an odd number as justified in \cite{CuMa11}. $4\times L\times M$ unknown coefficients exist  that are determined by a collocation method in which equations and boundary conditions
are imposed at the collocation points, according to the rules detailed in \cite{CuMa11}. Expansion orders  $L$ and $M$ are taken to ensure accuracy on the results: details of their values are provided in {the  Section 4.}

\subsection{Time-dependent schemes}
The governing equations \eqref{eqproblem1}--\eqref{eqproblem3} and their boundary conditions define a time-dependent problem for which we propose a temporal scheme based on a spectral  spatial discretization   analogous to that proposed in the previous section.
As before, expansion orders  $L$ and $M$ are such that they ensure accuracy on the results;  details on their values are given in the following section.
In order to integrate in time, we use a  third order multistep scheme. In particular,  we use a backward differentiation formula (BDF) that is  adapted for use with a variable time step. Details on the step adjustment are found in \cite{CuMa11}.
 BDFs are a particular case of multistep formulas which are  {\it implicit}. In \cite{CuMa11}, it is reported that instead of solving
the fully implicit scheme, a semi-implicit scheme is able to provide results with a similar  accuracy and fewer CPU time requirements, and this is the method we  employ to obtain the time-dependent solutions.

\section{Results {and discussion}}
Our study is focused on  the solutions displayed by this system at a fixed aspect ratio $\Gamma=2.166$. As the system is forced to transport more energy by increasing the $\Ra$ number, the conductive solution becomes unstable, and new convective solutions are observed.  The bifurcation point for this primary  event occurs  at $\Ra \sim 55$. Furthermore, beyond this point  a sequence of bifurcations occur when $\Ra$ increases, which is described below.

  \begin{figure}
    \hspace{-2cm}
   \includegraphics[width=17.5cm]{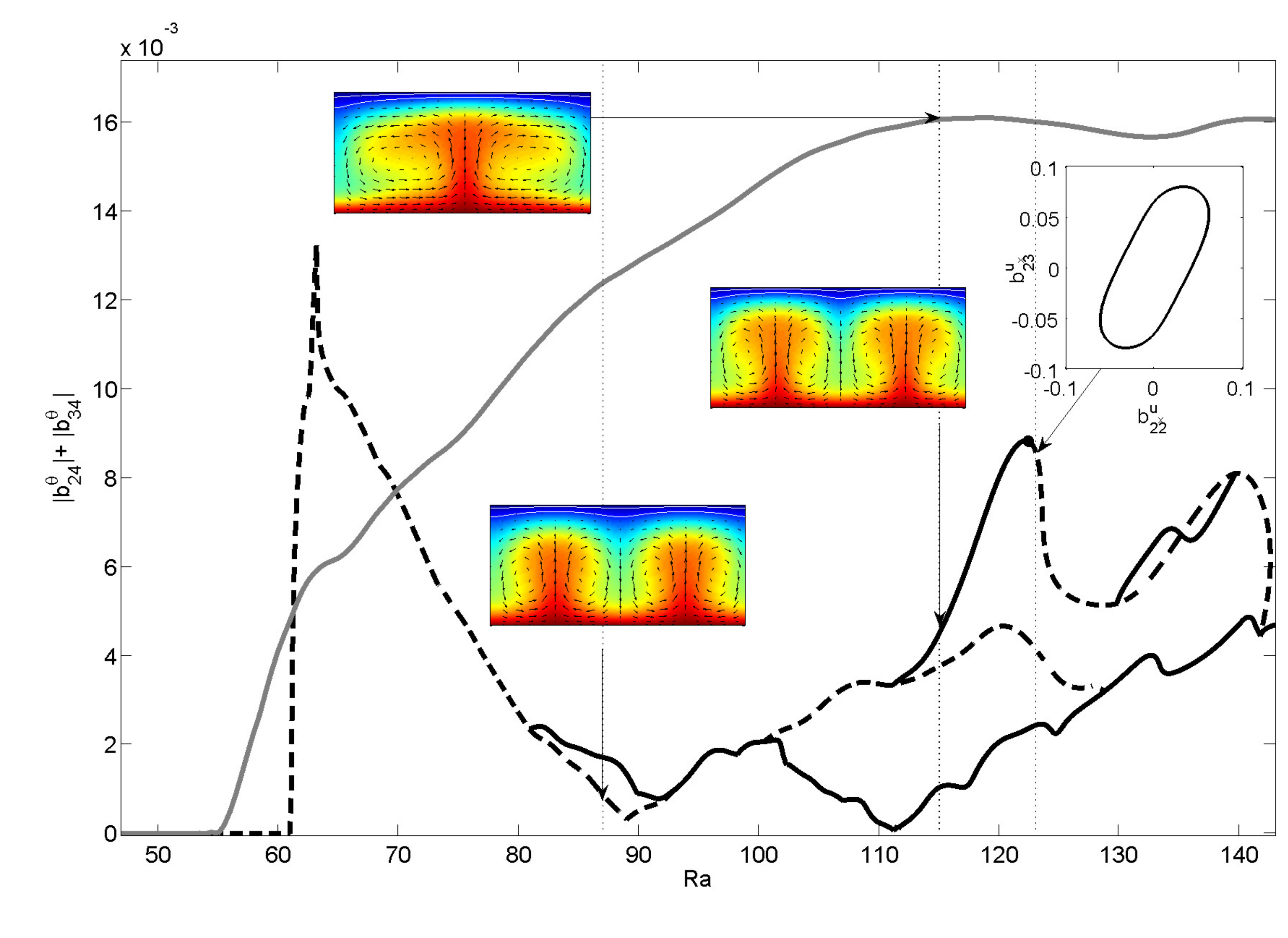}
    \caption{Bifurcation diagram  at $\Gamma=2.166$ for  the fluid under consideration in the range $\Ra\in [45,143]$. This
     is obtained by representing the amplitude $|b_{24}^\theta|+|b_{34}^\theta|$ versus the $\Ra$ number. Solid lines represent stable branches, while dashed lines stand for the unstable ones.} \label{bif2}
   \end{figure}

     \begin{figure}
     \centering
   \includegraphics[width=17.5cm,angle=90]{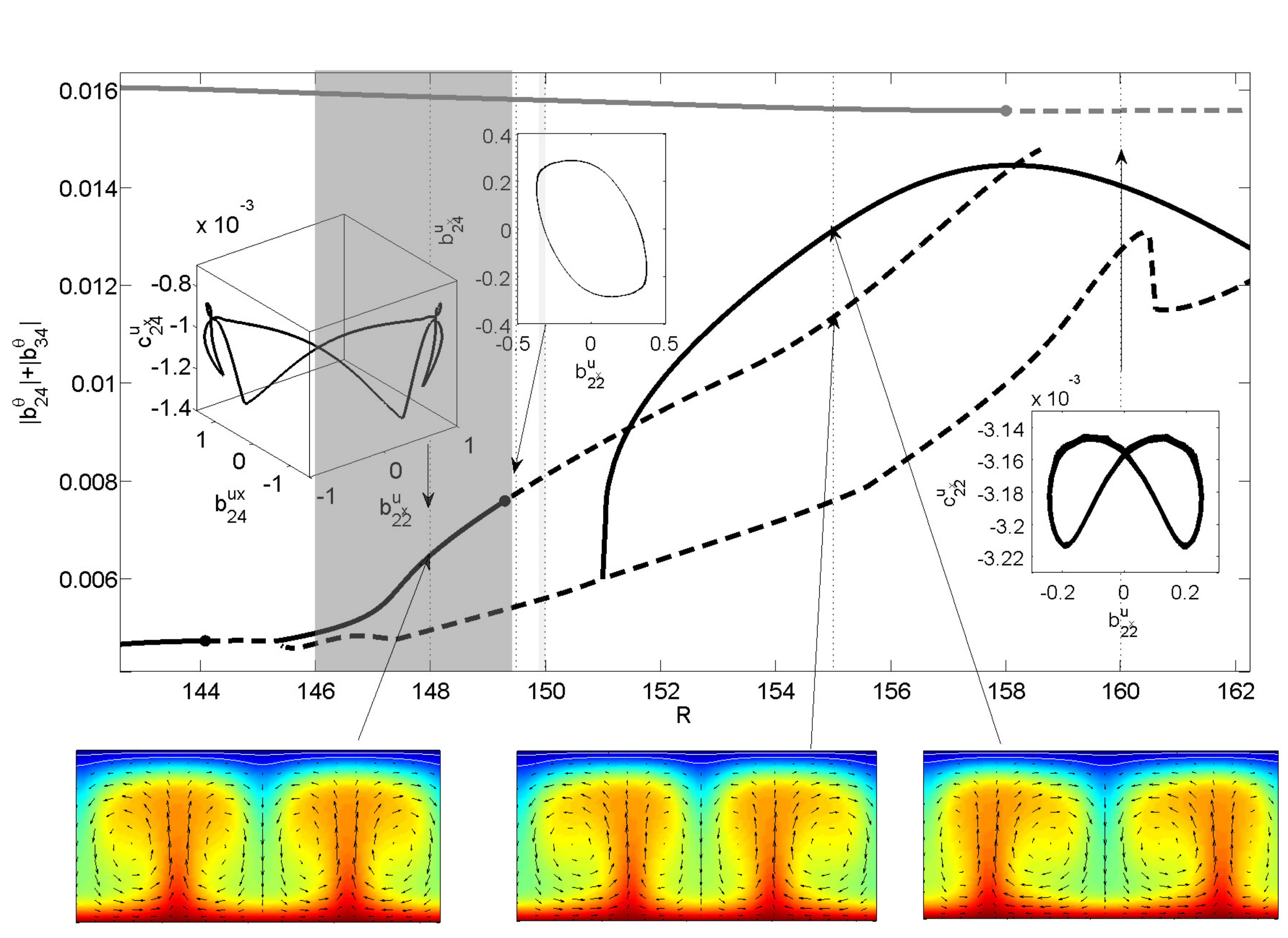}
    \caption{Bifurcation diagram  at $\Gamma=2.166$ for  the fluid under consideration in the range $\Ra\in [143,162]$. This
     is obtained by representing the amplitude $|b_{24}^\theta|+|b_{34}^\theta|$ versus the $\Ra$ number. Solid lines represent stable branches, while dashed lines stand for the unstable ones. Shaded regions limit the parameter values $\Ra$, between which the time-dependent solutions described in Figure \ref{Fsolplate} are observed.}\label{bif3}
   \end{figure}

     \begin{figure}
     \centering
   \includegraphics[width=7cm]{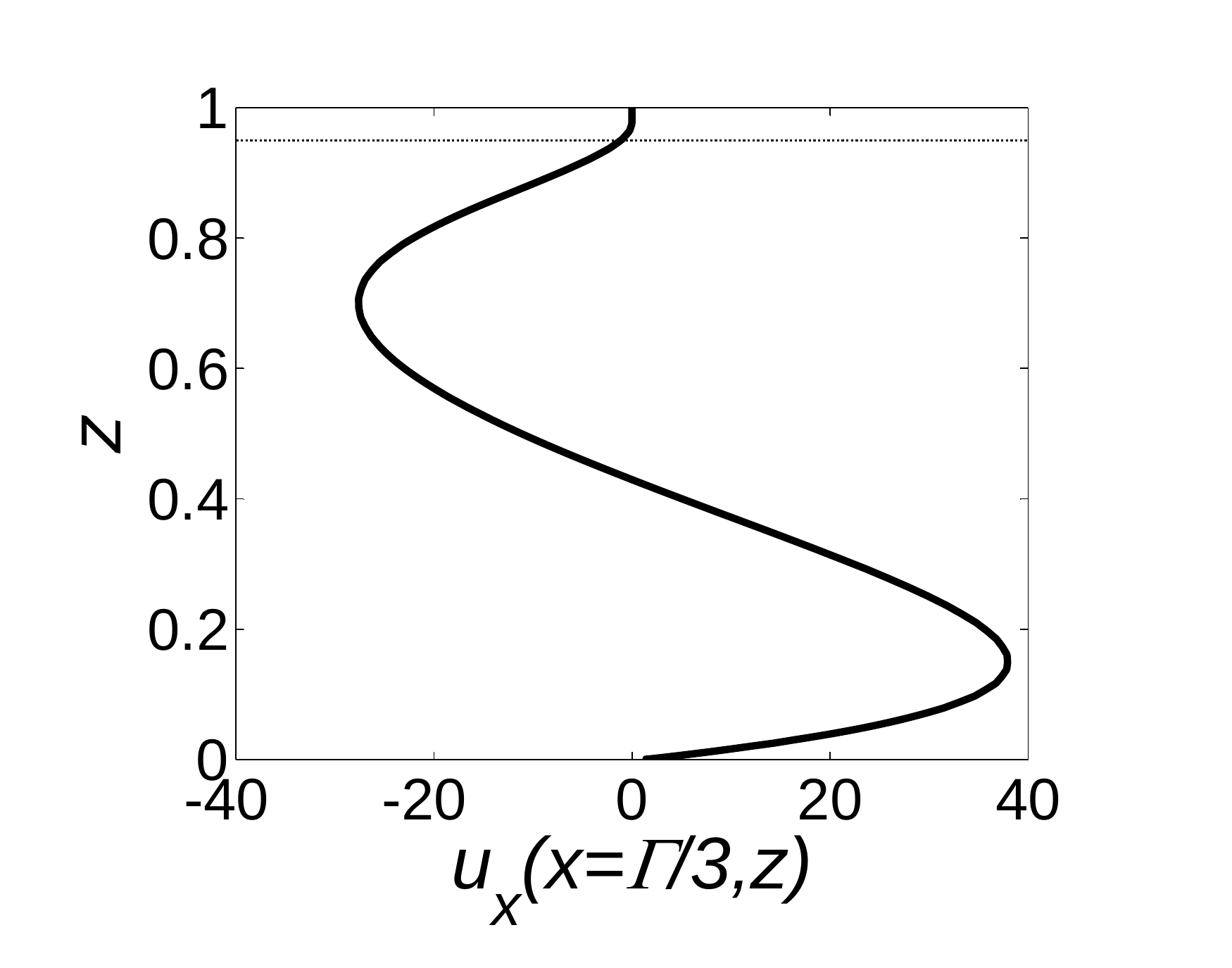}
    \caption{{Representation of $u_x$ versus $z$ for the stationary one plume pattern  obtained at  $\Gamma=2.166$ and $\Ra=115$. The  horizontal line highlights the stagnant upper lid.  }}\label{staglid}
   \end{figure}

     \begin{figure}
     \centering
    \subfigure[]{\includegraphics[width=11cm]{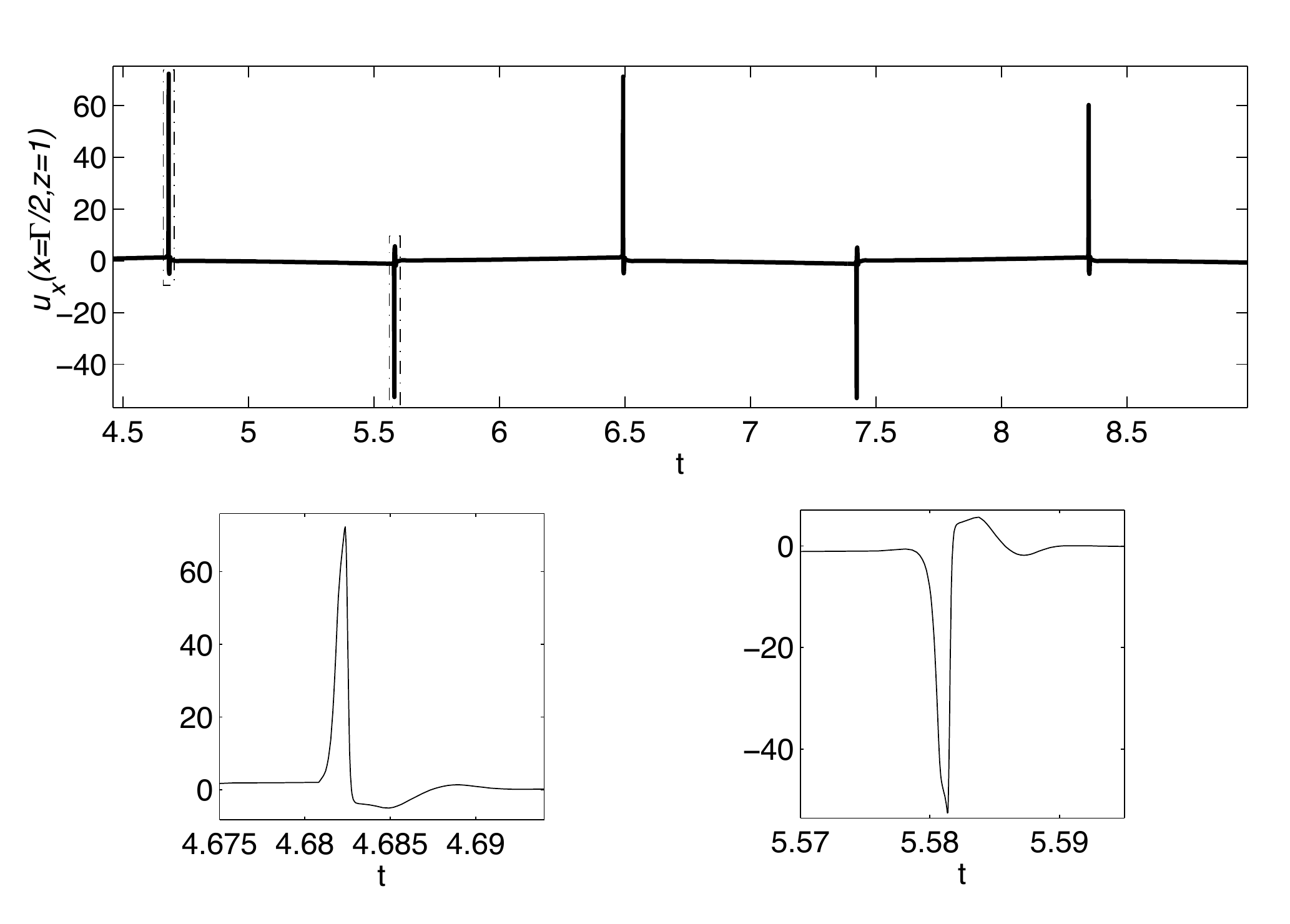}\label{Fsolplatea}}\\
    \subfigure[]{\includegraphics[width=5cm]{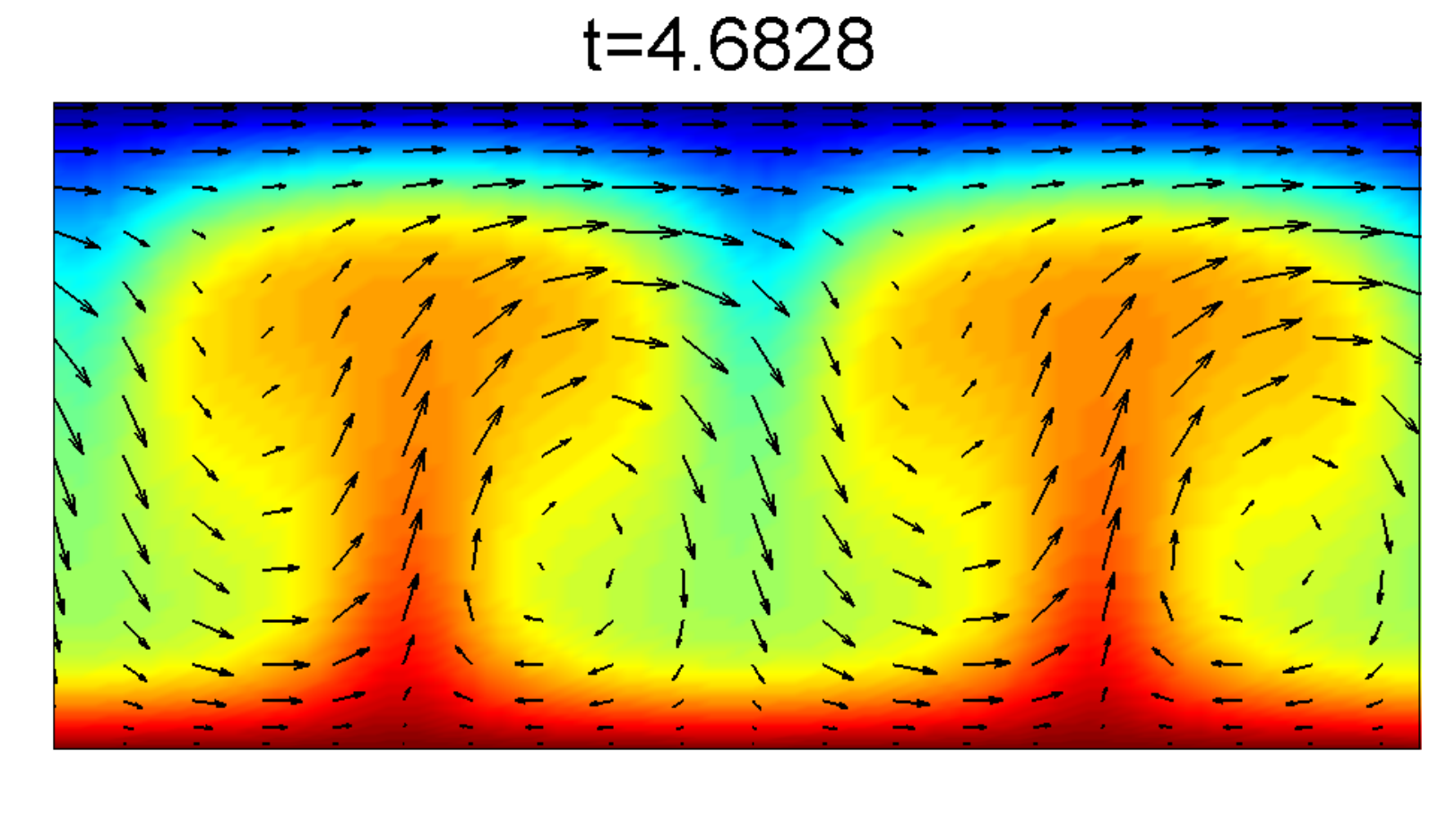}\includegraphics[width=5cm]{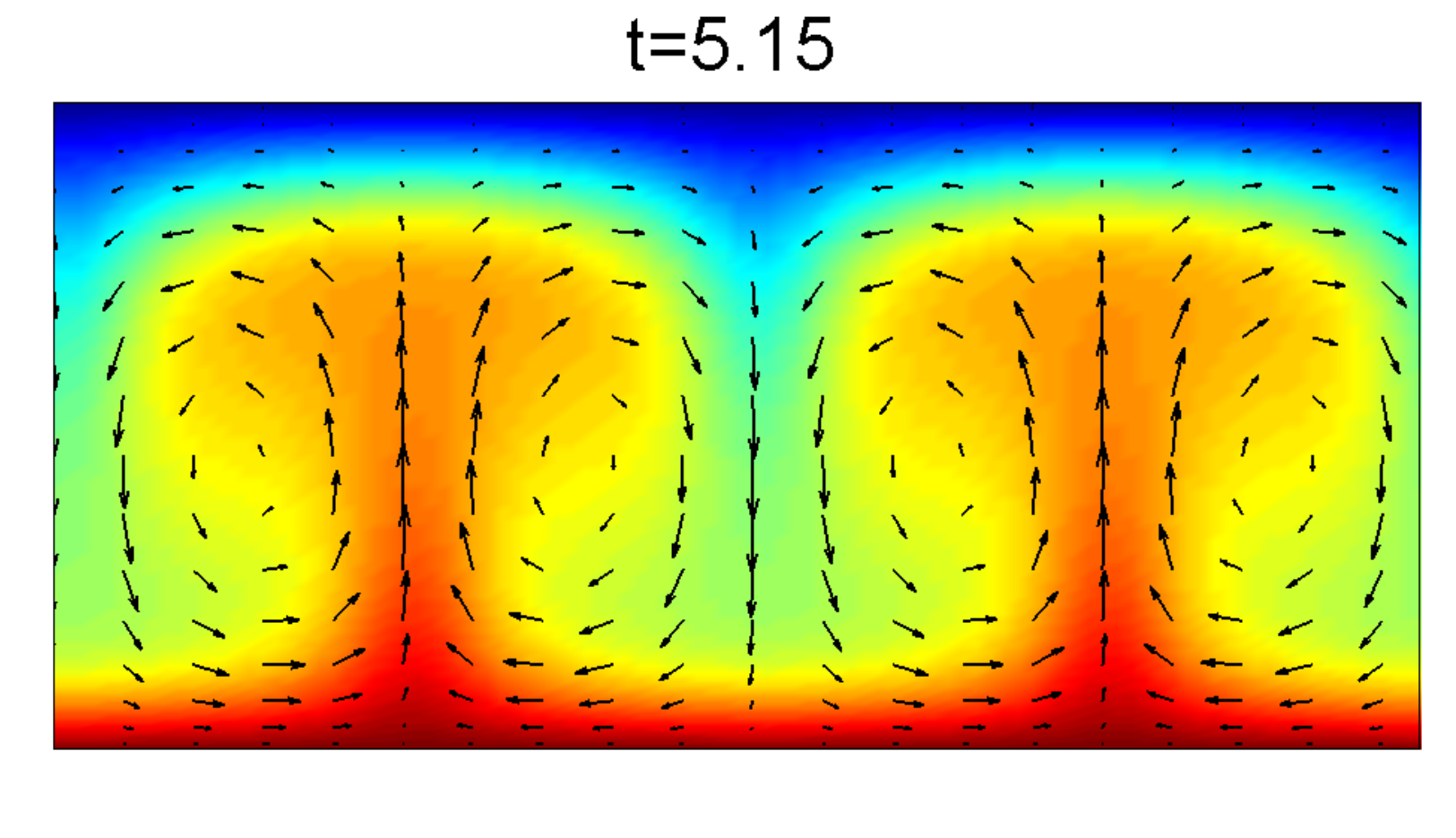}\includegraphics[width=5cm]{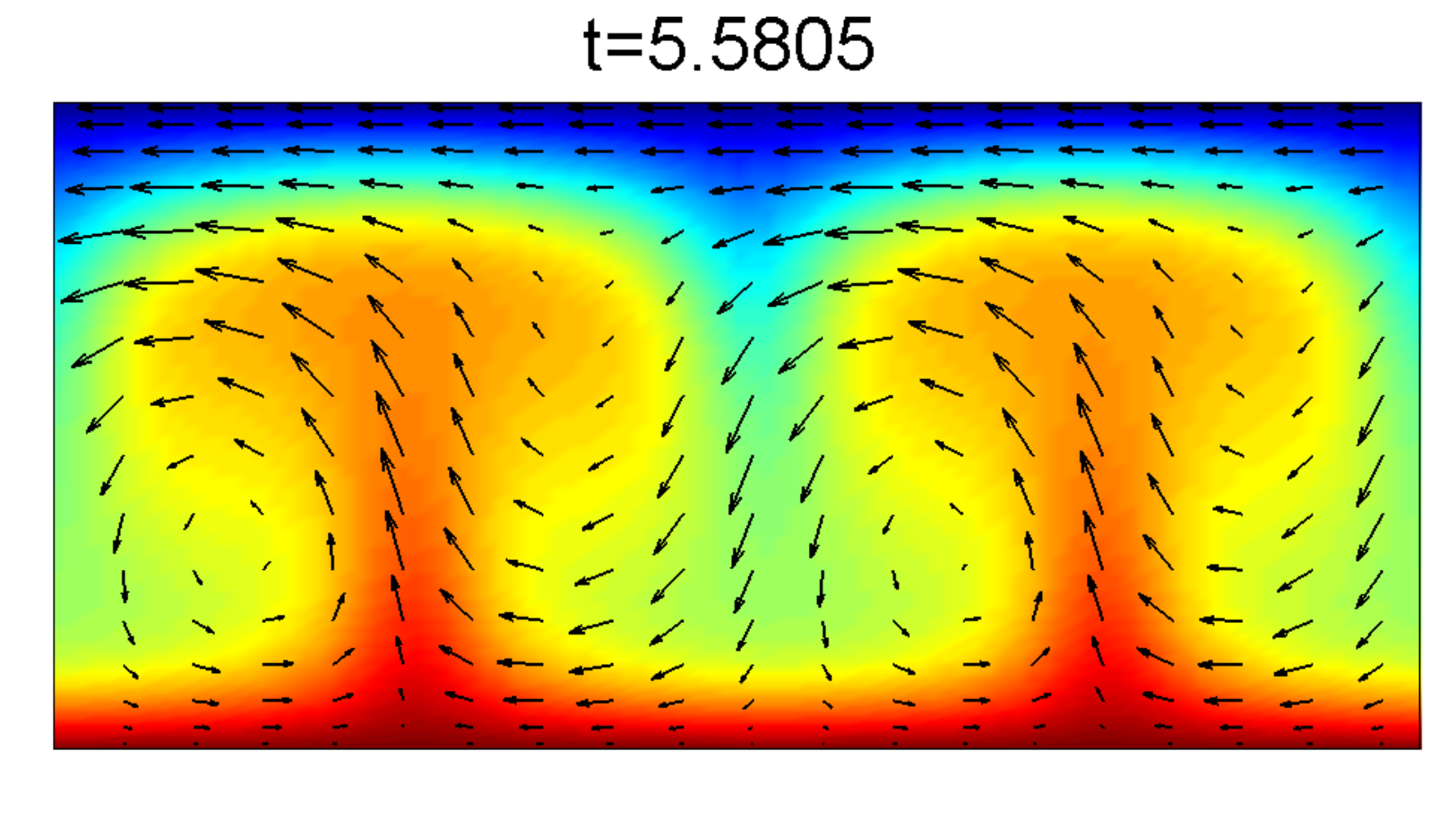}\label{Fsolplateb}}
    \subfigure[]{\includegraphics[width=5cm]{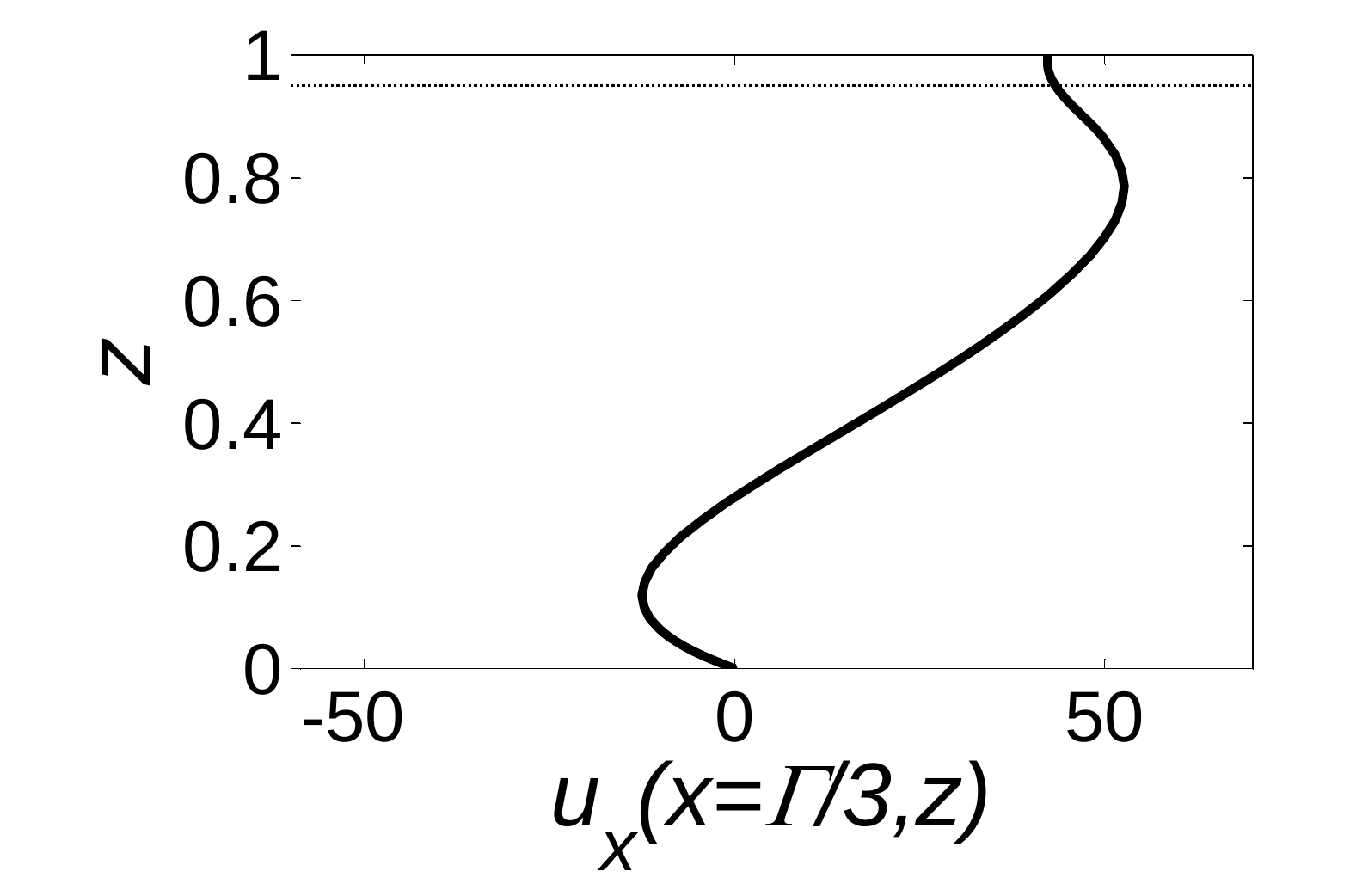}\includegraphics[width=5cm]{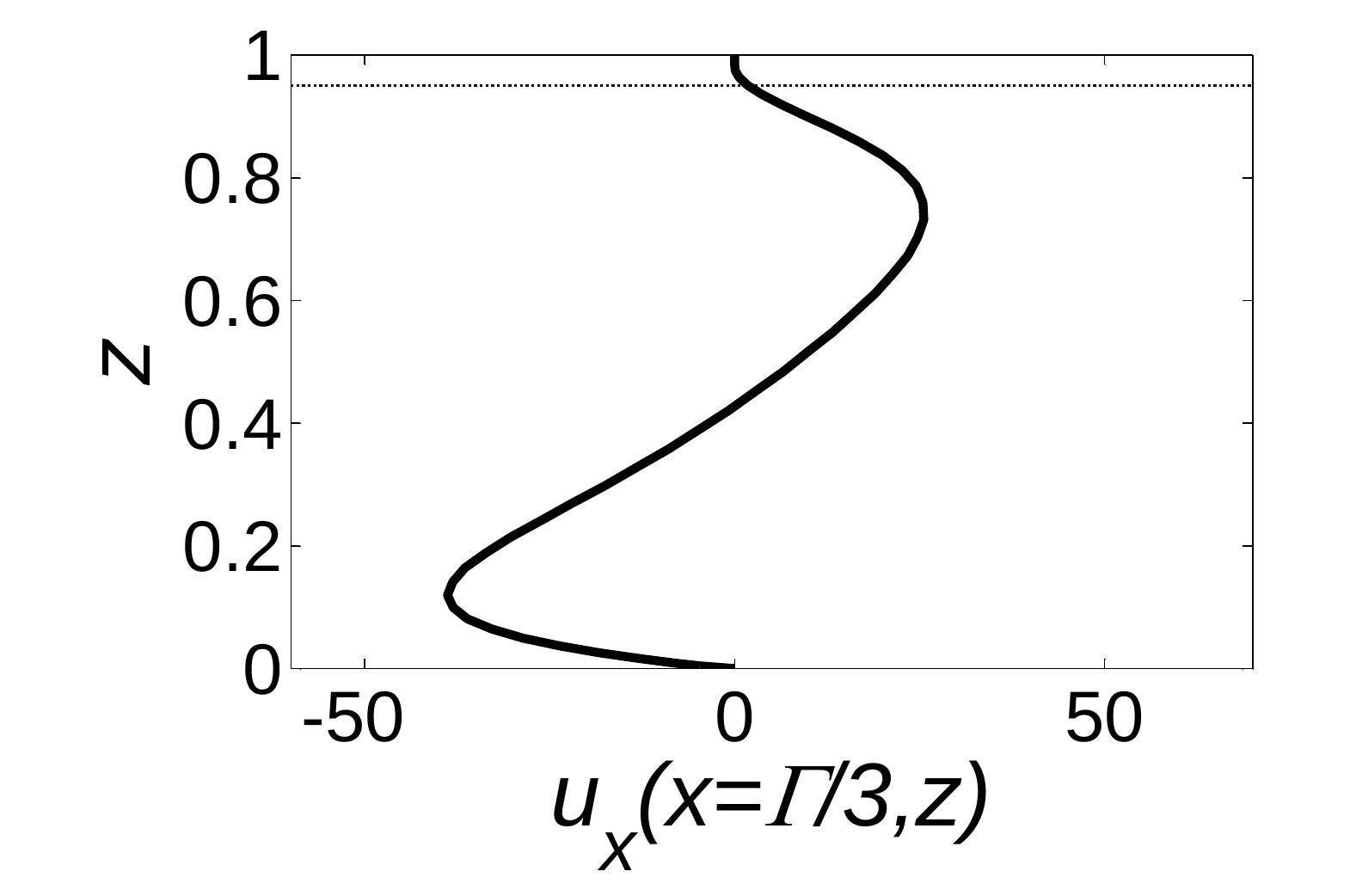}\includegraphics[width=5cm]{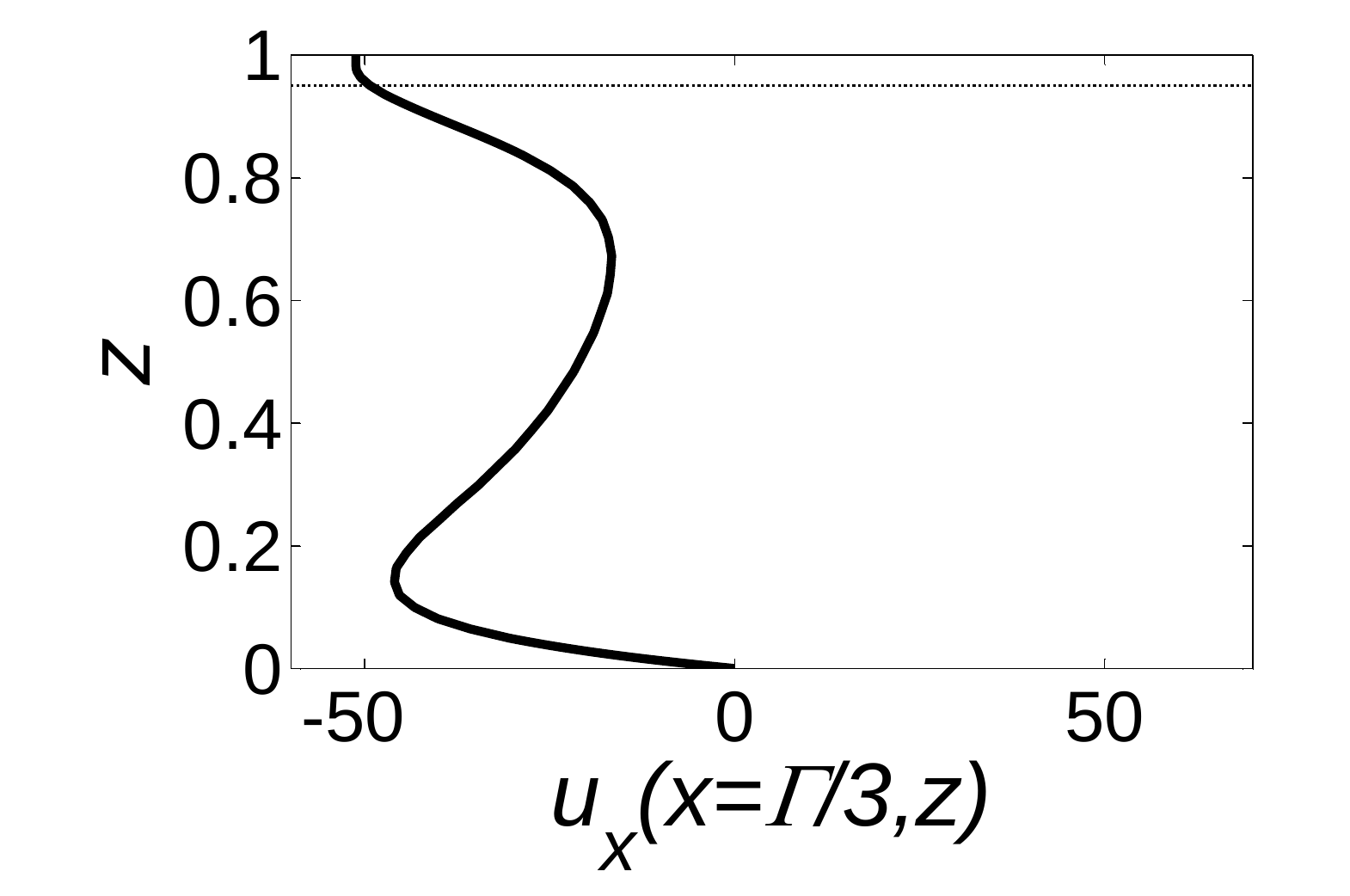}\label{Fsolplatec}}

    \caption{Time-dependent regime of a two-plume pattern at $\Gamma=2.166$ and $\Ra=148$ in the $\Ra$ range highlighted with the shaded region in Figure \ref{bif3}.   a) Time series of the horizontal component of the velocity at the surface point $(z=1,x=\Gamma/2)$ on $\Ra=148$.  Two zooms are represented  for the bursts at around times $t\sim4.7$ and $t\sim5.6$; b) spatial patterns during the bursts at  times $t=4.6828$ and $t=5.5805$ and in the quiescent state at $t=5.15$. {c) the horizontal component of the velocity $u_x$ versus the $z$-coordinate
    at fixed $x=\Gamma/3$ at  times   $t=4.6828$, $t=5.15$ and $t=5.5805$. The horizontal line highlights the moving upper lid that switches with a stagnant lid.}}\label{Fsolplate}
   \end{figure}

Figures \ref{bif2} and \ref{bif3} show the bifurcating branches captured at different $\Ra$ ranges. In the diagrams, the horizontal axis represents the $\Ra$ number, while on the vertical axis the system state is represented by a scalar  given by the sum of two coefficients in the expansion (\ref{eqexpansion3}) of a stationary solution:
\begin{align}\label{amplitude}
|b^{\theta}_{24}|+|b^{\theta}_{34}|.
\end{align}
This amplitude is related to  the energy in the temperature field.  The horizontal line at 0 corresponds to the conductive solution which is always a stationary solution of the system.
In these figures, stable branches are represented by solid lines, while unstable branches are shown with dashed lines. The lines in black correspond to solutions with periodicity $m=2$ and the lines in gray are for those with $m=1$.  The validity of these bifurcation diagrams   is decided by ensuring that for successive order expansions the amplitude values displayed on the vertical axis of the  bifurcation diagrams  are preserved. Most of the results reported in these diagrams are obtained with expansions $L\times M=47 \times 50$ although these are increased up to  $L\times M=50 \times 100$ when required (especially at high $\Ra$ numbers).

Figure \ref{bif2} is focused on the $\Ra$ interval  $\Ra\in [45, 143]$ and  reveals that several stable solutions are possible under the same physical conditions.
The patterns observed in the physical  variables, temperature and velocity, are plotted at different Rayleigh numbers.
For instance, at $\Ra=115$ the solution for the solid grey branch  is displayed,
which corresponds to a pattern with one plume. Two white lines indicate
the temperature contours at which the viscosity mostly decay. { Figure \ref{staglid} displays the horizontal component of the velocity versus the $z$-coordinate for this solution. The highlighted thin layer at the upper part confirms  the existence of a stagnant lid at the surface.}
At $\Ra=87$ a two-plume solution in an unstable branch  is represented.
  Several pitchfork bifurcations occur  from which stable branches emerge.
 At $\Ra=115$  the pattern of a two-plume stable solution is shown;  as in the one-plume solution,  it has a stagnant upper surface.  This stable branch undergoes a Hopf bifurcation at $\Ra\sim 123$, after which  time-dependent solutions are found. A projection on the expansion coefficients space of this time-dependent solution  is displayed at $\Ra\sim 123$. This solution consists of two plumes    each slightly oscillating around their axis  below a stagnant lid.
 After the Hopf bifurcation, the unstable branch undergoes a pitchfork bifurcation at $\Ra\sim 130$ and a stable branch emerges. This branch merges again with the unstable branch at $\Ra\sim 140$. The crossing of branches at
 $\Ra\sim 135$ is an effect of the  projection taken and does not represent any transcritical bifurcation.

 When the system is forced to transfer higher energy rates, further time-dependent solutions are observed. This is confirmed  in Figure \ref{bif3}, which describes a bifurcation diagram similar to the previous one but at higher Rayleigh numbers  $\Ra\in [143, 162]$.  The black branch bifurcates through a Hopf bifurcation at $\Ra\sim144$  towards time-dependent regimes that coexist  with the grey stable branch.
 Hopf bifurcations also occur for a different black stable branch  at $\Ra\sim 149.3$, as well as for the grey solid branch at $\Ra\sim 158$. Figure \ref{bif3} shows several stationary solutions:  at $\Ra\sim148$ it displays a two-plume structure obtained over the marked branch. This solution coexists  with a time-dependent solution  observed over the entire shaded region, of which we provide  a projection of the time series obtained at $\Ra=148$, {which is  explained in detail below.} An unstable stationary solution over this branch is displayed at $\Ra\sim155$. The pattern is asymmetric, with the plumes more  prominent  outwards.  At this $\Ra$ number, the structure of a stationary solution  on the stable branch also exhibits  asymmetry,  but with the heads  more prominent inwards. All the stationary solutions have a stagnant upper surface.
      \begin{figure}
     \centering
    \subfigure[]{\includegraphics[width=12.5cm]{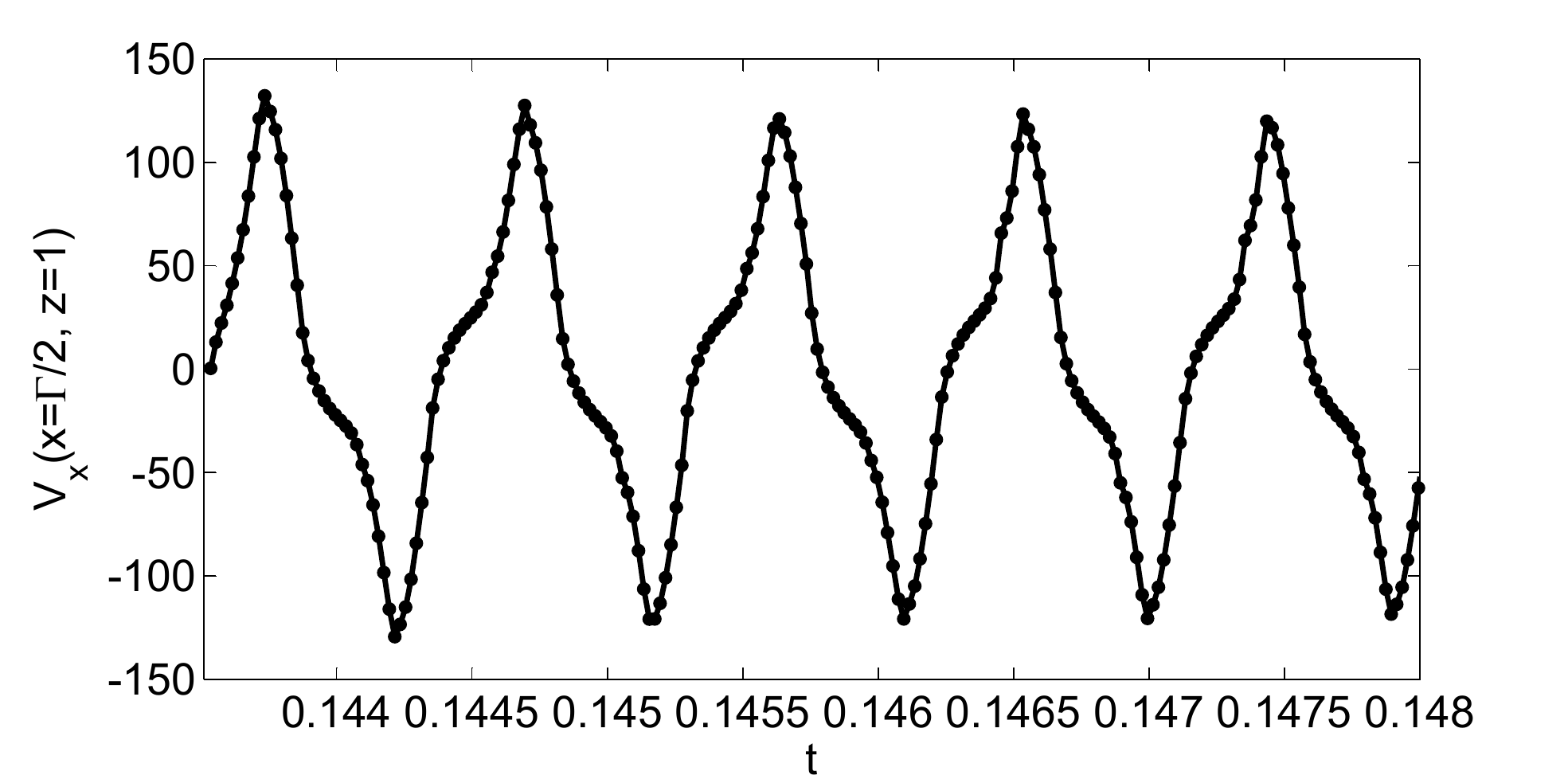}\label{tdoneplumea}}\\
    \subfigure[]{\includegraphics[width=5cm]{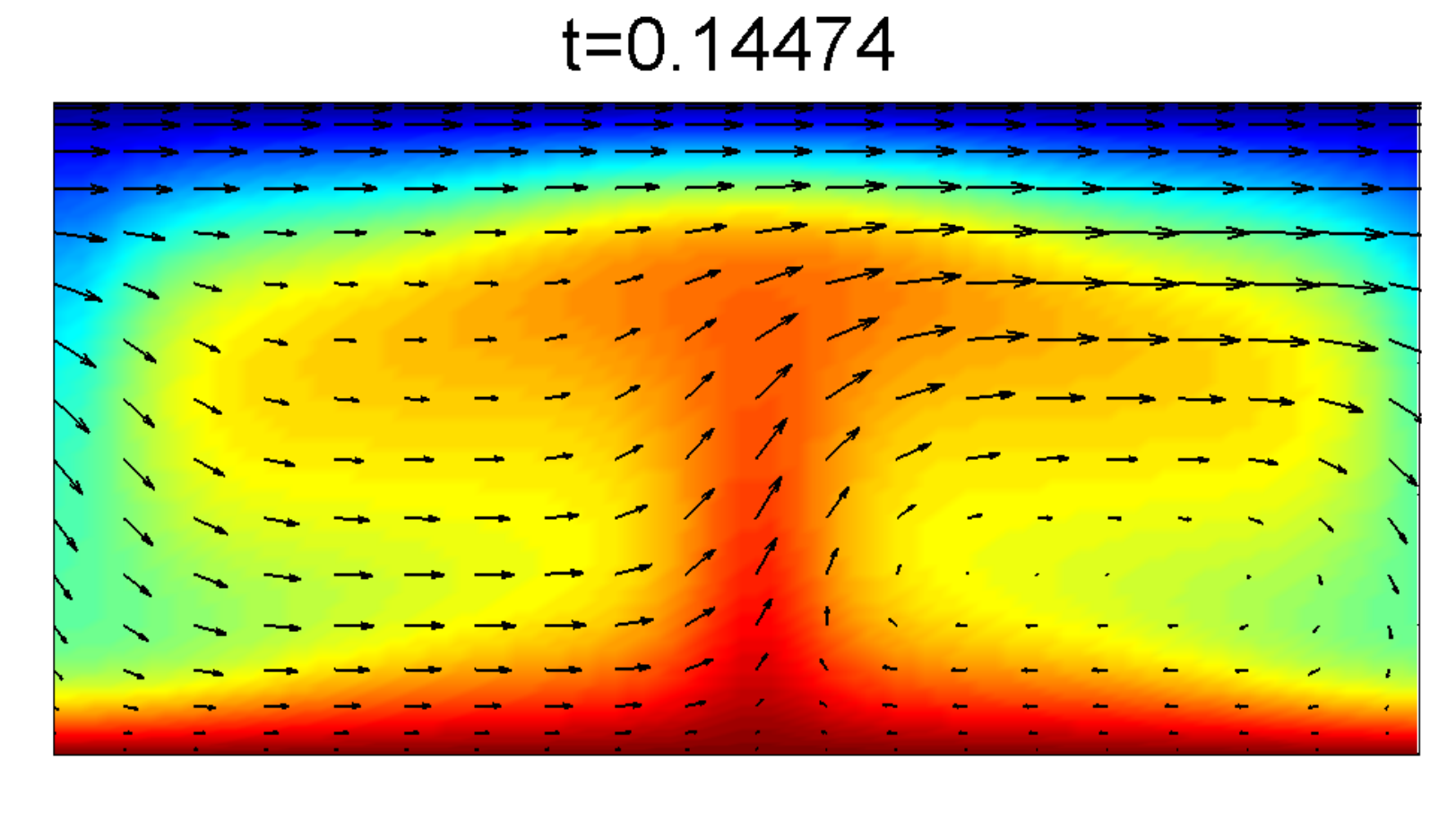}\includegraphics[width=5cm]{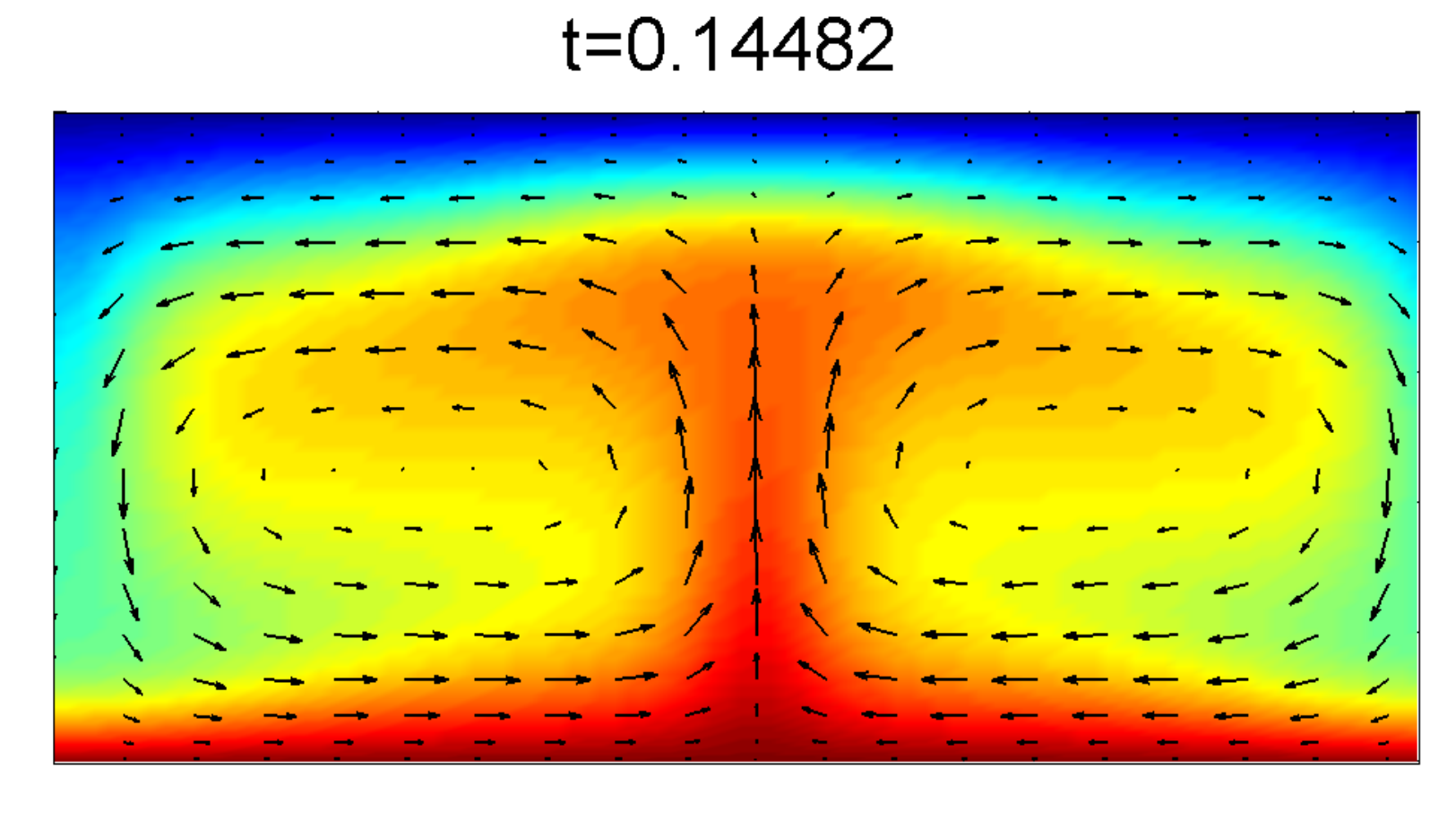}\includegraphics[width=5cm]{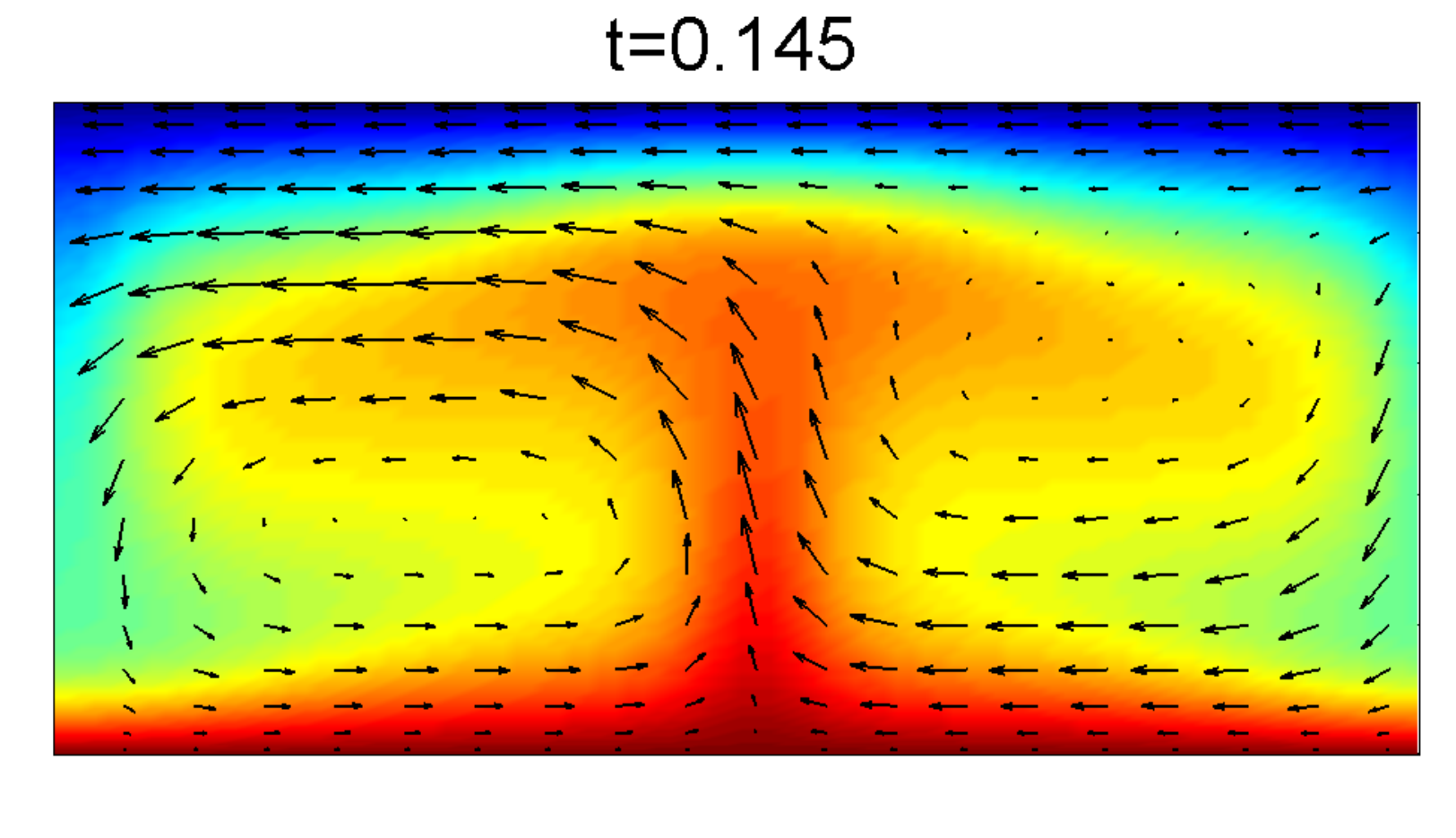}\label{tdoneplumeb}}
    \subfigure[]{\includegraphics[width=5cm]{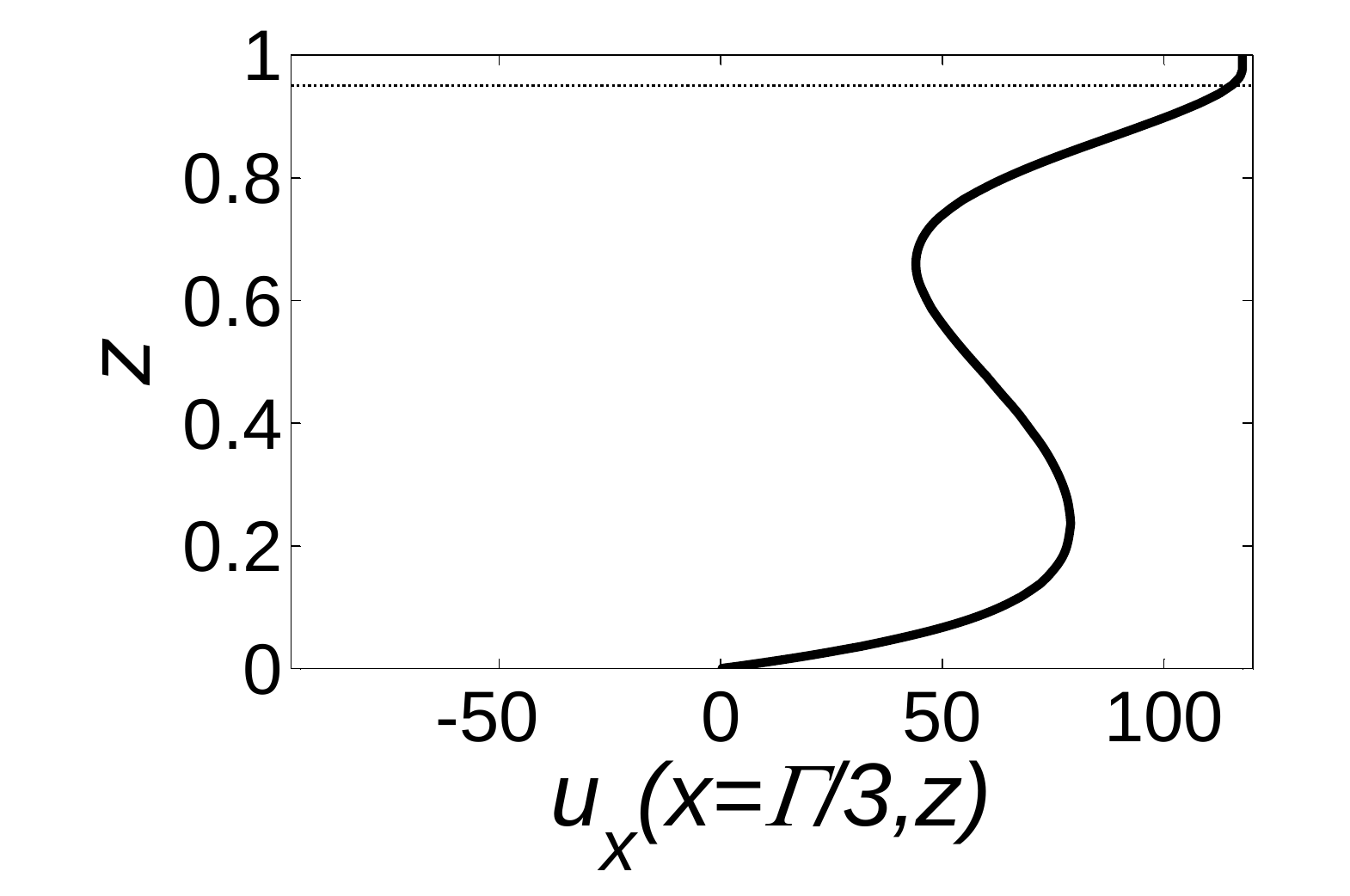}\includegraphics[width=5cm]{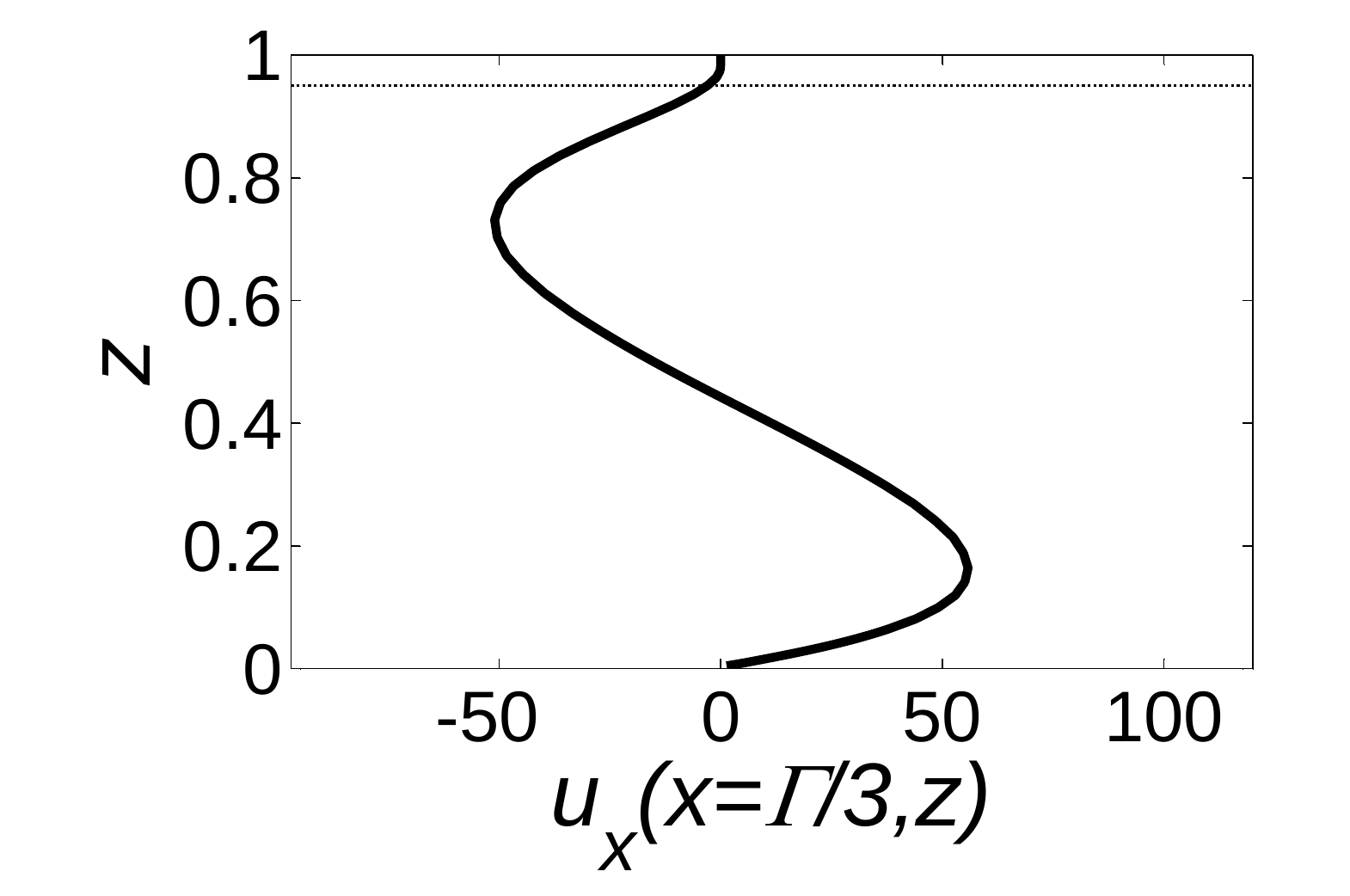}\includegraphics[width=5cm]{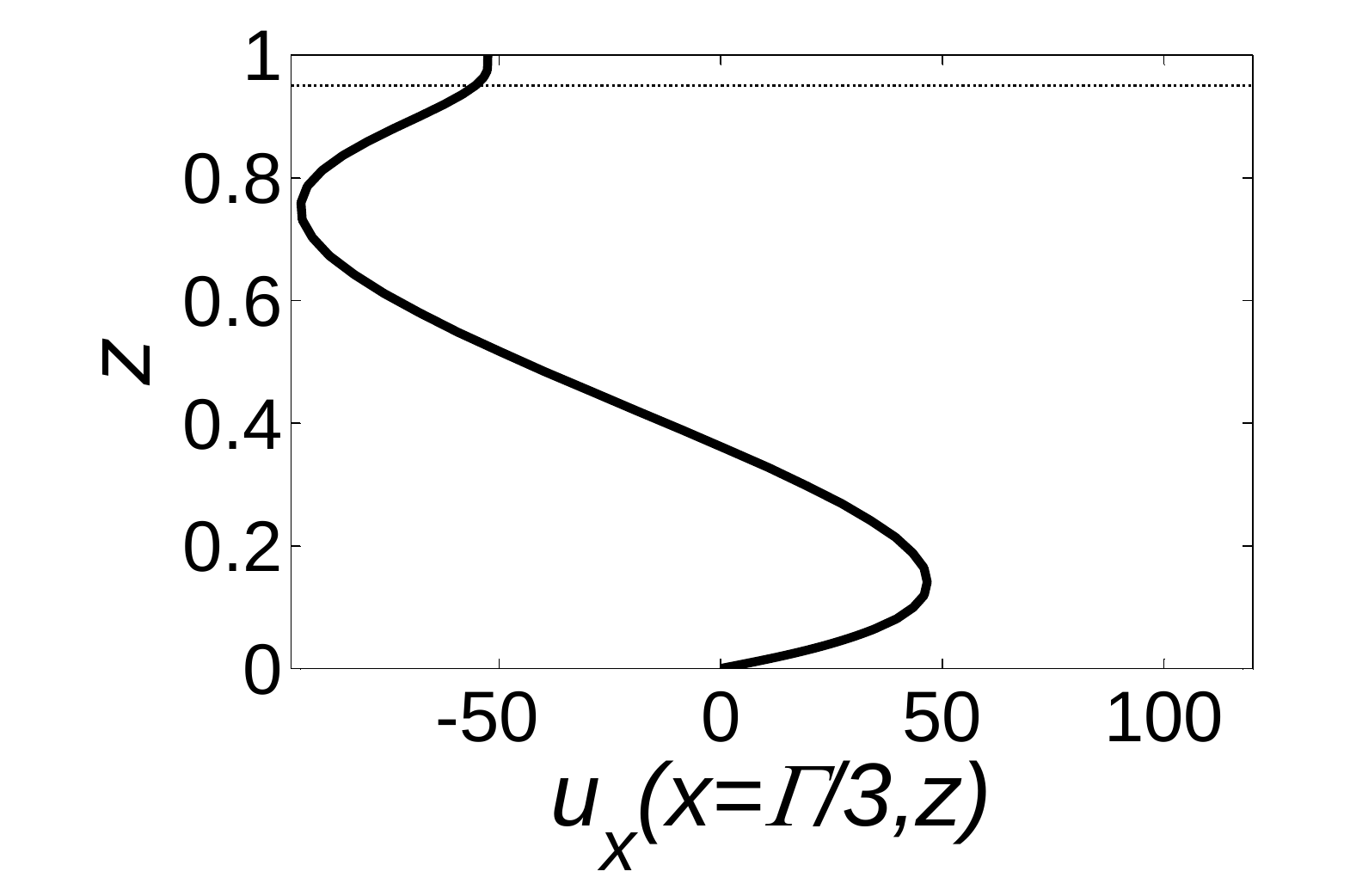}\label{tdoneplumec}}
    \caption{ Time-dependent regime of  a one-plume pattern at $\Gamma=2.166$ and $\Ra=160.$ a) Time series of the horizontal component of the velocity at the surface point $(z=1,x=\Gamma/2)$ at $\Ra=160$; b) spatial patterns  at  times $t=0.14474$ and $t=0.14482$ and  $t=0.145.$; {c) the horizontal component of the velocity $u_x$ versus the $z$-coordinate
    at fixed $x=\Gamma/3$ at  times $t=0.14474$, $t=0.14482$ and  $t=0.145$. The horizontal line highlights the moving upper lid.} }\label{tdoneplume}
   \end{figure}

 We next  describe the temporal evolution observed in   the shaded region of Figure \ref{bif3}.
 A summary of what is  obtained at $\Ra=148$ is given in Figure \ref{Fsolplate}. The system evolves as a limit cycle { in a type of evolution similar to that close to heteroclinic cycles, a typical object of systems with the O(2) symmetry \cite{AGH88,PoMa97}.  Over time, the system stays close to two quiescent states,  which are distinguished in  Figure \ref{Fsolplatea} by the zero upper velocity  for long periods. These states are interrupted by bursts in which energy at the upper surface is abruptly released.
 One of the two quiescent states is represented in the center panel  of  Figure \ref{Fsolplateb}. In these almost stationary  positions,} the system
presents two plumes that oscillate very slightly and have a stagnant lid at the upper surface.  The two states are distinguished  by the fact that the plumes are slightly shifted    along the horizontal direction.
{Figure  \ref{Fsolplatec} represents the horizontal component of the velocity $u_x$ versus the depth $z$ for a quiescent state at time $t=0.14482$. In the center panel one may observe the stagnant upper lid. The quasi-stationary regimes are connected by rapidly evolving transitions which release the energy very rapidly.
{The solution during these crises has the  interesting  characteristic that in these episodes it consists of "plate-like" convective styles.}
 By "plate-like" motion  we refer to the fact that the stagnant lid at the upper surface   drifts alternately towards the right or the left as a block. The first and third panels in Figure  \ref{Fsolplatec} show a moving upper layer  by displaying the horizontal component of the velocity $u_x$ versus the depth $z$.  A thin lid is observed which moves like  a rigid body without internal shear either to the right or to the left. We have verified that variations of the velocity within this thin layer  are below $0.05\%$,
  and for this reason it has the appearance of a moving plate.}
 In these short time intervals (short when compared to the duration of the quiescent states), a meandering jet develops simultaneously below the drifting surface, in which  sinking and upwards currents are observable (see Figure \ref{Fsolplateb}).

It is clear that the lateral boundary conditions (\emph{i.e.} the symmetry) are {important}  for this behaviour because they allow upper drift motions. Moreover, the shifted inactive states between which the alternancy appears are  possible only  in this scenario. { On the other hand,  in a recent work  by Ulvrov\'a et al. \cite{ULCRT12} the authors study a law similar to ours, although  symmetry effects are absent and  this type of transition is not reported.     Nevertheless, it should be noted  that symmetry is  not a sufficient condition for this kind of behavior:  for instance, previous results discussed in \cite{CM13}, also in the presence of the O(2) symmetry,    and in a similar setting  to ours, make no reference to  "plate-like" convection nor stagnant lids,  although symmetries do exert an influence on the described solutions. The main difference between the setting considered in this work and that presented in   \cite{CM13}  is that the transition of the viscosity with temperature is less  abrupt than in our study. Furthermore  results reported in \cite{CuMa11}, also in the presence of the O(2) symmetry and viscosity according to  exponential law, indicate the existence of a stagnant lid, although no  impact of the symmetry on the time evolution is found.

The bursting solutions  obtained in this study are justified in the framework of the symmetry presence, but this does not imply that episodic solutions cannot be obtained in other settings in which the symmetry is absent.  For instance, in the numerical work by  \cite{W93}, catastrophic events in Earth's mantle are reported. However, these bursting solutions  are rather different to ours insofar as they do not connect almost quasi-stationary states and the solutions are not linked to any plate-like behaviour. The work   by  \cite{KLG02} is connected to episodic plate reorganizations, although  the authors   obtain  time series in which transitions are not so dramatic as ours.
However, in their problem  they consider a square box with periodic boundary conditions, so they have the symmetry O(2)$\times$O(2). Unfortunately, they provide no discussion on the impact of the underlying symmetry on their solutions.

Our simulations indicate that the solution described in Fig.  \ref{Fsolplate} becomes non-attractive at $\Ra\sim 149.5$. }At this point,  an initial data  starting near this regime  evolves towards a periodic solution, a projection of which is shown in Figure \ref{bif3}. This motion consists of slightly asymmetrical plumes, each one rapidly vibrating around its  central vertical  axis. This is a time-dependent solution in which the surface fluid remains stagnant.
At $\Ra\sim 160$, Figure \ref{bif3} shows the projection  on the coefficient space of a time series obtained for the asymptotic time-dependent regime of one plume. Figure \ref{tdoneplume}
shows  the time  evolution in detail. The system evolves in a periodic motion in which the upper surface  drifts alternately towards the right and the left, which is confirmed by a  time series of the horizontal velocity component
 at the surface. Long quiescent states between the drift motions are no longer observed, only a continuously oscillating motion. { In  Figure \ref{tdoneplumeb}, snapshots  of the temperature and velocity fields obtained at times $t=0.14474, 0.14482, 0.145$ show the plate-like motion.  Figure \ref{tdoneplumec} enforces this vision by displaying   the horizontal component of the velocity $u_x$ versus the depth $z$ at the same selected times. A thin upper lid which moves consistently is observed.} As in the previously described plate motion,  a meandering jet develops below the drifting surface in which  sinking and upwards currents are observable (see Figure \ref{tdoneplumeb}).
These time-dependent solutions are obtained with expansions  $L\times M=47 \times 50$.

\section{Conclusions}
{In this paper we  address the subject  of a convecting fluid in which viscosity depends on temperature. We  examine  a dependency which models an abrupt change in the viscosity   in a gap around a temperature of transition.} We  explore the space of solutions at a fixed aspect ratio by means of bifurcation diagrams and time-dependent numerical simulations. We find time-dependent convection in which the symmetry plays an important role.
 In particular, we describe  limit cycles and time periodic solutions which are similar to others found in several contexts in the literature (see \cite{AGH88,GH88,PoMa97}) in  the presence of the O(2) symmetry.

 {The time evolution during the limit cycles presents two peculiarities: first of all, they are bursting solutions that release energy abruptly in time and secondly plate-like convection  is observed during the bursts. Additionally, time-periodic solutions are found that have a similar plate-like dynamic with a smoother time evolution.
 No plate-like dynamics have hitherto been observed in this type of convection problem. For  viscosity dependencies according to  the Arrhenius law, or its approach by means of an exponential law, no temporal transitions between  stagnant lids and drifting lids have been reported. Recent studies by Ulvrov\'a et al. \cite{ULCRT12}, who use a law similar to ours, do not  report  this type of transitions either, although symmetry effects  are not considered in their  study.

Forsyth and Uyeda \cite{FU75} propose that plate-like motion is produced by  sinking slabs that  pull the plates in the subduction process.
The results reported in our study are obtained for constant density within the Boussinesq approximation, and provide convection examples  of moving plates  that coexist with subsurface upwards and downwards meandering jets, but without a proper subduction.
  Obviously, these examples  do not rule out  the existence of subduction in the Earth, but rather propose a role played by the  symmetry which
  can be particularly illustrative for understanding convective styles of the Earth prior to subduction, or that  of other planetary bodies. }


 \section*{Acknowledgements}

We thank  CESGA  for computing facilities. This research is supported by the Spanish Ministry of Science under grant MTM2011-26696 and MINECO: ICMAT Severo Ochoa project SEV-2011-0087.

\bibliographystyle{plain}
\bibliography{local}

\begin{thebibliography}{10}

\bibitem{AGH88}
D.~Armbruster, J.~Guckenheimer, and P.~Holmes.
\newblock Heteroclinic cycles and modulated travelling waves in systems with
  {O}(2) symmetry.
\newblock {\em Physica D}, 29(257-282), 1988.

\bibitem{ben1900}
H.~Benard.
\newblock Les tourbillons cellulaires dans une nappe liquide author(s): Benard,
  h. source: Rev. gen. sci. pures appl. volume: 11 pages: 1261-1271 published:
  1900 times cited: 589 (from web of science).
\newblock {\em Rev. Gen. Sci. Pures Appl.}, 11:1261--1271, 1900.

\bibitem{B03}
D.~Bercovici.
\newblock The generation of plate tectonics from mantle convection.
\newblock {\em Earth and Planetary Science Letters}, 205(107-121), 2003.

\bibitem{B75}
F.~H. Busse.
\newblock Pattern of convection in spherical shells.
\newblock {\em J. Fluid Mech.}, 72:65--85, 1975.

\bibitem{B82}
F.~H. Busse and N.~Riahi.
\newblock Pattern of convection in spherical shells {II}.
\newblock {\em J. Fluid Mech.}, 123:283--391, 1982.

\bibitem{C75}
P.~Chossat.
\newblock Bifurcation and stability of convective flows in a rotating or not
  rotating spherical shell.
\newblock {\em SIAM Journal on Applied Mathematics}, 37:624--647, 1975.

\bibitem{CK91}
J.~D. Crawford and E.~Knobloch.
\newblock Symmetry and symmetry-breaking bifurcations in fluid dynamics.
\newblock {\em Annu. Rev. Fluid Mech.}, 23(341-387), 1991.

\bibitem{CM13}
J.~Curbelo and A.~M. Mancho.
\newblock Bifurcations and dynamics in convection with temperature-dependent
  viscosity under the presence of the {O}(2) symmetry.
\newblock {\em Physical Review E}, 2013.

\bibitem{CuMa11}
J.~Curbelo and A.~M. Mancho.
\newblock Spectral numerical schemes for time-dependent convection with
  viscosity dependent on temperature.
\newblock {\em Communications in Nonlinear Science and Numerical Simulations},
  19(2), 2014.

\bibitem{Dav01}
G.F. Davies.
\newblock {\em Dynamic Earth. {P}lates, Plumes and Mantle convection}.
\newblock Cambridge University Press, Cambridge, England, 2001.

\bibitem{PoMa97}
S.~P. Dawson and A.~M. Mancho.
\newblock Collections of heteroclinic cycles in the {K}uramoto-{S}ivashinsky
  equation.
\newblock {\em Physica D: Nonlinear Phenomena}, 100(3-4):231--256, 1997.

\bibitem{Dubu}
F.~Dubuffet, D.~A. Yuan, and E.~S.~G. Rainey.
\newblock Controlling thermal chaos in the mantle by positive feedback from
  radiative thermal conductivity.
\newblock {\em Nonlinear Proc. Geophys.}, 9(1-13), 2002.

\bibitem{FU75}
D.~Forsyth and S.~Uyeda.
\newblock On the relative importance of the driving forces of plate motion.
\newblock {\em Geophys. J. R Astr. Soc.}, 43:163--200, 1975.

\bibitem{GS82}
M.~Golubitsky and D.G. Schaeffer.
\newblock Bifurcation with {O}(3) symmetry including applications to the benard
  problem.
\newblock {\em Communs. Pure. Appl. Math.}, 35:81--11, 1982.

\bibitem{GH88}
J.~Guckenheimer and P.~Holmes.
\newblock Structurally stable heteroclinic cycles.
\newblock {\em Math. Proc. Cambridge Philos. Soc.}, 103(189-192), 1988.

\bibitem{hof}
A.~M. Hofmeister and D.~Yuen.
\newblock Critical phenomena in thermal conductivity: Implications for lower
  mantle dynamics.
\newblock {\em Journal of Geodynamics}, 44:186--199, 2007.

\bibitem{HHM02}
S.~Hoyas, H.~Herrero, and A.~M. Mancho.
\newblock Bifurcation diversity of dynamic thermocapillary liquid layers.
\newblock {\em Physical Review E}, 66(5):057301, 2002.

\bibitem{HMHGD05}
S.~Hoyas, A.~M. Mancho, H.~Herrero, N.~Garnier, and A.~Chiffaudel.
\newblock B{\'e}nard--{M}arangoni convection in a differentially heated
  cylindrical cavity.
\newblock {\em Physics of Fluids}, 17:054104, 2005.

\bibitem{IG84}
E.~Ihrig and M.~Golubitsky.
\newblock Pattern selection with {O}(3) symmetry.
\newblock {\em Physica D}, 12:1--33, 1984.

\bibitem{jarvis}
G.~T. Jarvis and D.~P. Mckenzie.
\newblock Convection in a compressible fluid with infinite prandtl number.
\newblock {\em J. Fluid Mech.}, 96(03):515--583, 1980.

\bibitem{KLG02}
S.~D. King, J.~P. Lowman, and C.~W. Gable.
\newblock Episodic tectonic plate reorganizations driven by mantle convection.
\newblock {\em Earth and Planetary Science Letters}, 203:83--91, 2002.

\bibitem{zhong}
X.~Liu and S.~Zhong.
\newblock Analyses of marginal stability, heat transfer and boundary layer
  properties for thermal convection in a compressible fluid with infinite
  prandtl number.
\newblock {\em Geophys. J. Int.}, 194:125--144, 2013.

\bibitem{MW91}
P.~Machetel and P.~Weber.
\newblock Intermittent layered convection in a model mantle with an endothermic
  phase change at 670 km.
\newblock {\em Nature}, 350:55--57, 1991.

\bibitem{MBH97}
A.~M. Mancho, H.~Herrero, and J.~Burguete.
\newblock Primary instabilities in convective cells due to nonuniform heating.
\newblock {\em Physical Review E}, 56(3):2916, 1997.

\bibitem{MS95}
L.~N. Moresi and V.~S. Solomatov.
\newblock Numerical investigation of 2{D} convection with extremely large
  viscosity variations.
\newblock {\em Physics of Fluids}, 7(9):2154--2162, 1995.

\bibitem{NMH07}
M.~C. Navarro, A.~M. Mancho, and H.~Herrero.
\newblock Instabilities in buoyant flows under localized heating.
\newblock {\em Chaos: An Interdisciplinary Journal of Nonlinear Science},
  17:023105, 2007.

\bibitem{R82}
D.~Rand.
\newblock Dynamics and symmetry: predictions for modulated waves in rotating
  fluids.
\newblock {\em Arch. Ration. Mech. Anal.}, 79(1):1--38, 1982.

\bibitem{TH98}
R.Trompert and U.~Hansen.
\newblock Mantle convection simulations with reologies that generate plate-like
  behaviour.
\newblock {\em Nature}, 395(686-689), 1998.

\bibitem{R73}
D.~Ruelle.
\newblock Bifurcations in the presence of a symmetry group.
\newblock {\em Arch. Ration. Mech. Anal.}, 51:136--152, 1973.

\bibitem{S04}
V.~S. Solomatov.
\newblock Initiation of subduction by small-scale convection.
\newblock {\em J. Geophys. Res}, 109:B01412, 2004.

\bibitem{SM96}
V.~S. Solomatov and L.~N. Moresi.
\newblock Stagnant lid convection on {V}enus.
\newblock {\em J. Geophys. Res}, 101:4737--4753, 1996.

\bibitem{SM97}
V.~S. Solomatov and L.~N. Moresi.
\newblock Three regimes of mantle convection with non-newtonian viscosity and
  stagnant lid convection on the terrestrial planets.
\newblock {\em Geophysical Research Letters}, 24(15):1907--1910, 1997.

\bibitem{Setal92}
S.~C. Solomon, S.~E. Smrekar, D.~L. Bindschadler, R.~E. Grimm, W.~M~. Kaula,
  R.~J. Phillips, R.~S. Saunders, G.~Schubert, S.~W. Squyres, and E.~R. Stofan.
\newblock Venus tectonics: An overview of {M}agellan observations.
\newblock {\em J. Geophys. Res}, 97(13199-132555), 1992.

\bibitem{T98}
P.~J. Tackley.
\newblock Self-consistent generation of tectonic plates in three dimensional
  mantle convection.
\newblock {\em Earth and Planetary Science Letters}, 157:9--22, 1998.

\bibitem{ULCRT12}
M.~Ulvrov\'a, S.~Labrosse, N.~Coltice, P.~Raback, and P.J. Tackley.
\newblock Numerical modelling of convection interacting with a melting and
  solidification front: Application to the thermal evolution of the basal magma
  ocean.
\newblock {\em Physics of the Earth and Planetary Interiors}, 206-207:51--66,
  2012.

\bibitem{W93}
S.~A. Weinstein.
\newblock Catastrophic overturn of the earth's mantle driven by multiple phase
  changes and internal heat generation.
\newblock {\em Geophysical Research Letters}, 20(321-324), 1993.

\bibitem{Yana}
T.~K.~B. Yanagawa, M.~Nakada, and D.~A. Yuan.
\newblock The influence of lattice thermal conductivity on thermal convection
  with strongly temperature-dependent viscosity.
\newblock {\em Earth Space Sci.}, 57(15-28), 2005.

\end{thebibliography}

\newpage

\newpage

\newpage
\newpage

\end{document}